\documentclass[a4paper,11pt]{article}
\pdfoutput=1 
\usepackage[english]{babel}
\usepackage{microtype,xspace}
\usepackage[utf8]{inputenc}	

\usepackage{jheppub} 
\usepackage{bm,amssymb,amsmath,slashed,graphicx,multirow,soul,mathtools,xspace,array,epsfig,color,bm}
\usepackage{siunitx}
\usepackage{bm} 
\usepackage{float}
\usepackage{feynmf}
\allowdisplaybreaks
\usepackage{bbold}
\usepackage[normalem]{ulem} 

\usepackage[usenames,dvipsnames]{xcolor}
\definecolor{red}{rgb}{0.6,.0706,.1373}
\definecolor{blue}{rgb}{0,0.396,0.741}

\usepackage{subcaption}
\usepackage{pdflscape}

\usepackage[T1]{fontenc}

\title{\boldmath The remarkable role of the vector-like quark doublet in the Cabibbo angle and $W$-mass anomalies}

\author[1,2]{Benedetta Belfatto,}
\author[3]{Sokratis Trifinopoulos}

\affiliation[1]{SISSA International School for Advanced Studies,
  Via Bonomea 265, 34136, Trieste, Italy}
\affiliation[2]{INFN - Sezione di Trieste,
Via Bonomea 265, 34136, Trieste, Italy}
\affiliation[3]{Center  for  Theoretical  Physics,  Massachusetts  Institute  of  Technology,  Cambridge,  MA  02139,  USA}

\abstract{
The combination of precise determinations of $V_{us}$ and $V_{ud}$ hints towards a violation of the CKM first row unitarity at about $3\sigma$ level. Conversely, the recent measurement of the $W$-boson 
mass by the CDF Collaboration exhibits significant tension with the SM prediction, intensifying the conflict that may arise in models addressing the unitarity violation. We demonstrate that one vector-like $SU(2)_L$ doublet with mass of a few TeV mixing with light quarks and the top quark can simultaneously account for the two anomalies, without conflicting with flavour-changing phenomena and electroweak observables. Moreover, another tension in the value of the Cabibbo angle is reported also at $3\sigma$ level, between two determinations of $V_{us}$ obtained from semi-leptonic $K\ell3$ and leptonic $K\mu 2$ kaon decays. We show that the vector-like doublet
can be at the origin of this discrepancy and the substantial positive shift in the $W$-boson mass. This unique feature of the vector-like quark doublet may render it a crucial puzzle piece in new physics scenarios addressing the CKM unitarity problem. The model can be potentially probed in future colliders.}

\begin{document}

\maketitle

\flushbottom
\newpage

\section{Introduction}

The Standard Model (SM) of particle physics has unambiguously proven to be extremely resilient against a vast plethora of high-precision measurements designed to tests its limits. 
Among the last remaining puzzles are the the so-called Cabibbo angle anomalies, the tensions between three different determinations of the Cabibbo angle. Meanwhile, the $W$-boson mass $m_W$ as measured by the CDF Collaboration significantly departs from the SM prediction~\cite{CDF:2022hxs}. 
This set of anomalous observations is quite recent and it is important to scrutinize their validity and consistency both at experimental as well as theoretical level.

As regards the Cabibbo angle $\theta_C$, recent calculations of short-distance radiative corrections
in $\beta$-decays led to an improved determination 
of $\vert V_{ud} \vert=\cos\theta_C$. 
Meanwhile, experimental data on kaon decays combined with recent theoretical and lattice computations
provide a precise determination of $\vert V_{us} \vert=\sin\theta_C$ and $\vert V_{us}/V_{ud}\vert =\tan\theta_C$.
These three determinations
are in tension between each other and two anomalies
arise.
The first Cabibbo angle anomaly (CAA1) can be identified as the deficit in the CKM first row unitarity relation when confronting the
value of $|V_{ud}|$ from $\beta$-decays with the value of $|V_{us}|$ from kaon decays.
The second Cabibbo angle anomaly (CAA2) stems from the 
two different measurements of $|V_{us}|$, i.e. 
the direct determination of $\vert V_{us} \vert$ provided by semi-leptonic $K\ell 3$ kaon decays, 
and the determination of the ratio
$\vert V_{us}/V_{ud}\vert $ obtained from leptonic $K\mu 2$ and $\pi\mu 2$ decay rates. 
The significance of each anomaly individually amounts to approximately $3 \sigma$.  

These anomalies, if confirmed with future data, would indicate the presence of physics beyond the SM. Various models that include mediators at the TeV scale have been suggested as possible solutions to CAA1 \cite{Belfatto:2019swo,Grossman:2019bzp,Coutinho:2019aiy,Cheung:2020vqm,Crivellin:2020lzu,Endo:2020tkb,Belfatto:2021jhf,Branco:2021vhs,Capdevila:2020rrl,Crivellin:2020ebi,Kirk:2020wdk,Manzari:2020eum,Alok:2020jod,Crivellin:2020oup,Crivellin:2020klg,Crivellin:2021njn,Marzocca:2021azj,Fischer:2021sqw,Botella:2021uxz,Crivellin:2022rhw,Babu:2022jbn}. However, in general, explanations for the CKM unitarity can be in conflict with the 
new experimental development for $m_W$, which points towards an enhancement with respect to the SM. 
For example, before the CDF-II result it was suggested in ref. \cite{Belfatto:2019swo} that a solution to CAA1 by means of modification of the Fermi constant $G_F$ with respect to the muon decay constant $G_\mu$ also imply a deficit of $m_W$. Reversely, in ref. \cite{Cirigliano:2022qdm}, it is shown that the models that predict a positive shift in the $W$ mass may also predict a huge violation of CKM unitarity, much larger than the one indicated by the current anomaly.

Among the proposed mediators for the anomalies the case of vector-like quarks is particularly interesting. Their mixing with the first generation SM quarks can consistently resolve the CAA1 at tree-level~\cite{Belfatto:2019swo,Belfatto:2021jhf,Branco:2021vhs,Crivellin:2022rhw,Crivellin:2022rhw,Babu:2022jbn}, while at one-loop level they can generate a positive contribution to $m_W$ via their coupling to the top \cite{Crivellin:2022fdf,Cao:2022mif,Babu:2022jbn}. 
Intriguingly, it was shown in ref.~\cite{Belfatto:2021jhf} that a specific type of vector-like quark charged as a doublet of $SU(2)_L$ and coupling predominantly to the up and strange quarks can also accommodate CAA2. However, stringent constraints from flavour-changing phenomena forbid the simultaneous explanation of both Cabibbo angle anomalies. To this end, at least two generations of vector-like doublets are required.

In this work we show that taking into consideration the CDF-II result, the vector-like quark doublet emerges as a favoured candidate mediator for the Cabibbo angle anomalies. In particular, the solution to the CKM unitarity deficit based on this field is compatible with a sufficiently large positive shift in $m_W$. Equivalently, a combined solution can be achieved for the tension between $K\ell3$ and $K\mu2/\pi\mu2$ determinations of $V_{us}$ and the $m_W$ discrepancy. We study the phenomenology for each scenario and chart the relevant parameter space that satisfies all the other low-energy constraints. Additionally, we assess the sensitivity of future experiments on the predictions of the model.

The paper is organized as follows. First we summarize the current situation of the Cabibbo angle anomalies in section \ref{sec:VusVud}. In section \ref{sec:doublet} we present the model of the vector-like quark doublet and provide explicitly the mixing matrices with the SM quarks. We  also present the anomalous observables in terms of the model parameters and list all the other relevant constraints. Subsequently, in section \ref{sec:discussion} we present the results of the phenomenological analysis evaluating the improvement over the SM. We also discuss other possible one-particle scenarios with respect to a combined explanation, laying special emphasis on the non-trivial case of the vector-like quark singlets.
Finally, in section \ref{sec:conclusion} we conclude and discuss briefly the future prospects. 

\section{Status of Cabibbo angle anomalies}
\label{sec:VusVud}

The SM predicts the unitarity of $V_\text{CKM}$, which for the first row implies
\begin{align}
\label{eq:uns}
& |V_{ud}|^2+|V_{us}|^2+|V_{ub}|^2=1~.
\end{align}
Since the contribution of
$|V_{ub}|^2\sim 1.6\times 10^{-5}$ is too small, the condition in eq. \eqref{eq:uns} reduces to unitarity of Cabibbo mixing.
In particular, three different kind of information can be extracted on the Cabibbo angle:
$|V_{us}|_A=\sin\theta_C$, $|V_{ud}|_B=\cos\theta_C$, $|V_{us}/V_{ud}|_C=\tan\theta_C$. 
In this section we list the three independent determinations 
for the $|V_{us}|$ and $|V_{ud}|$ CKM matrix elements that drive the anomalies.
\begin{itemize}
    \item {\bf Determination A:} One precise determination of $\vert V_{us} \vert $ stems from
semi-leptonic $K\ell3$ decays $K \to \pi \ell \nu$
($K_L e3$, $K_S e3$, $K^{\pm}e3$, $K^{\pm} \mu3$, $K_L \mu3$, $K_S \mu3$).
Recently, a reanalysis of electromagnetic radiative corrections was performed \cite{Seng:2021nar,Seng:2022wcw}, obtaining results in agreement with previous chiral perturbation theory calculations \cite{Cirigliano:2008wn} 
but with reduced uncertainties.
After including updated values of experimental inputs, phase-space factors, radiative and isospin-breaking corrections,
the result is $f_+(0)|V_{us}|=0.21634(38)$ \cite{Seng:2022wcw}.
Using the average of 4-flavour lattice QCD calculations
for the vector form factor $f_+(0)=0.9698(17)$ as reported by FLAG 2021 \cite{FlavourLatticeAveragingGroupFLAG:2021npn} gives
\begin{align}
\label{eq:det_A}
\vert V_{us} \vert_\text{A} = 0.22308(55)~.
\end{align}
The element $|V_{us}|$ can also be determined from semi-leptonic hyperon decays 
$|V_{us}|=0.2250(27)$ \cite{Cabibbo:2003ea}
and from hadronic $\tau$ decays $|V_{us}|=0.2221(13)$ \cite{HFLAV:2019otj},
which however present quite large uncertainties and therefore are not included in the present analysis.

    \item {\bf Determination B:} The ratio $|V_{us}/V_{ud}|$ can be independently determined from
the ratio of the kaon and pion leptonic 
decay rates $K\mu 2$ and $\pi\mu2$, i.e. $K\rightarrow \mu\nu(\gamma)$ and  $\pi\rightarrow \mu\nu(\gamma)$ \cite{Marciano:2004uf} which after including electroweak radiative corrections \cite{Cirigliano:2011tm,Giusti:2017dmp,DiCarlo:2019thl}
yields $f_{K}|V_{ud}|/(f_{\pi}|V_{us}|)=0.27600(37)$ \cite{KLOEKLOE-2:2014tsu,Moulson:2017ive,Workman:2022ynf}. 
Using the 4-flavour average lattice QCD calculations for the decay constants ratio $f_{K}/f_{\pi}=1.1932(21)$ \cite{FlavourLatticeAveragingGroupFLAG:2021npn},
the result is 
\begin{align}
\label{eq:det_B}
\qquad |V_{us}|/|V_{ud}|_B =0.23131(51)~.
\end{align}

    \item  {\bf Determination C:} $\vert V_{ud} \vert $ can be obtained from $\beta$ decays.
The most precise determination is obtained from super-allowed $0^+$-- $0^+$ nuclear $\beta$ decays, 
pure Fermi transitions which are
sensitive only to the vector coupling $G_V=G_F \vert V_{ud} \vert$.
The master formula is \cite{Hardy:2014qxa,Hardy:2020qwl}
\begin{align}
\label{Vud-super}
& \vert V_{ud} \vert _{0^+-0^+} ^2 = \frac{K }{ 2 G_F^2 \mathcal{F} t\, (1+ \Delta^V_R)}  
=\frac{2984.432(3) }{ \mathcal{F} t\, (1+ \Delta^V_R)}~,
\end{align}
where $K = 2\pi^3 \ln 2/m_e^5 = 8120.27624(1) \times 10^{-10}$~s/GeV$^4$,
$G_F=G_\mu  = 1.1663787(6) \times 10^{-5}$~GeV$^{-2}$ is the Fermi constant determined 
from the muon decay \cite{Tishchenko:2012ie}, 
 $\Delta^V_R$ is the transition-independent short-distance radiative correction,
$\mathcal{F} t=ft(1+\delta_R')(1+\delta_{\text{NS}}-\delta_C)$ is the ``corrected" $\mathcal{F} t$-value obtained from
the $ft$-value (which depends on the transition energy, the half-life of the parent state and the relevant branching ratio) after including the transition-dependent part of radiative corrections ($\delta_R'$, $\delta_{\text{NS}}$) 
and isospin-symmetry-breaking correction $\delta_C$.
The $\mathcal{F} t$ value was recently updated in ref. \cite{Hardy:2020qwl}, averaging the corrected $ft$-values of $15$ super-allowed $0^+$-- $0^+$ nuclear transitions and obtaining $\mathcal{F} t = 3072.24(1.85)\, \text{s}$. 
Compared with the previous average, the new result is almost unchanged in the central value 
but the uncertainty is increased by a factor of $2.6$ due to 
new contributions in nuclear structure corrections $\delta_{\text{NS}}$ \cite{Seng:2018qru,Gorchtein:2018fxl}, 
which now dominate the uncertainty. The short-distance radiative correction $\Delta^V_R$ was recently calculated with reduced theory uncertainty,
and found to be
$\Delta_R = 0.02467(22)$ \cite{Seng:2018qru,Seng:2018yzq}, 
which 
is significantly larger than the previous determination $\Delta^V_R =0.02361(38)$ \cite{Marciano:2005ec}.
This shift was confirmed by other recent studies \cite{Czarnecki:2019mwq,Seng:2020wjq,Hayen:2020cxh,Shiells:2020fqp}.
We use the result $\Delta^V_R = 0.02467(22)$ \cite{Seng:2018qru,Seng:2018yzq}.
Then, eq. \eqref{Vud-super} yields $|V_{ud}|_{0^+-0^+}=0.97367(31)$.

One can determine $\vert V_{ud} \vert$ also from free neutron $\beta$-decay.
The master formula gives (for review see for example refs. \cite{Gonzalez-Alonso:2018omy,Dubbers:2021wqv})
\begin{align}
\label{neutron}
& \vert V_{ud} \vert_n^2 = \frac{K/ \ln 2 }{ G_F^2  \mathcal{F}_n \tau_n \, (1+3g_A^2)  (1+ \Delta^V_R)} 
=  \frac{5024.5(6)~{\rm s}}{\tau_n (1+3g_A^2)(1+ \Delta^V_R)}~,
\end{align}
where $\mathcal{F}_n = f_n(1+\delta'_R)$ with $f_n= 1.6887(2)$ and $\delta'_R = 0.014902(2)$ is
the long-distance QED correction \cite{Towner:2010zz}, $\tau_n$ the neutron lifetime and $g_{A}=G_{A}/G_{V}$, where $G_A$ is the axial-vector coupling. 
The neutron lifetime $\tau_n$ can be obtained from bottle experiments
counting survived ultra-cold neutrons stored in traps.
The average of eight results \cite{UCNt:2021pcg,Ezhov:2014tna,Pattie:2017vsj,Serebrov:2017bzo,Arzumanov:2015tea,Steyerl:2012zz,Pichlmaier:2010zz,Serebrov:2004zf} (with
rescaled uncertainty)
reads $\tau_n=878.4\pm 0.5$~s.
\footnote{The neutron lifetime $\tau_n$ can be experimentally obtained with two different methods, namely bottle experiments and beam experiments, which count protons produced in $\beta$-decay.
However, there is a ~$4\sigma$ tension between the results of the two methods. The average of beam experiments \cite{Yue:2013qrc,Byrne:1996zz} gives $\tau_n^\text{beam}=888.0\pm 2.0$~s. The average including
bottle and beam experiments with rescaled uncertainty would give $878.6\pm 0.6$~s (see also footnote \ref{foot-neut}).}
Regarding the axial-vector coupling, 
the quoted average is $g_{A}=-1.2754\pm 0.0013$ with inflated uncertainty because of the tension
between different results, which combined with bottle lifetime in eq. \eqref{neutron} would give $|V_{ud}|_{n,\text{PDG}}=0.97436(88)$. 
However, 
the latest experiments measuring 
parity-violating $\beta$-asymmetry parameter $A$ from polarized  
neutrons~\cite{Mund:2012fq,UCNA:2017obv,Markisch:2018ndu}
have produced the most precise results, in very good agreement between each other, 
yielding an average $g_A=-1.27624(50)$. They show some tension with old results 
and with one of the recent results obtained measuring the electron-antineutrino angular correlation ($a$ coefficient) 
from the recoil spectrum of protons, 
$g_A=-1.2677(28)$~\cite{Beck:2019xye} by the aSPECT experiment. Nevertheless there is agreement with the other recent measurement of the $a$ coefficient by the aCORN 
experiment $g_A=-1.2796(62)$~\cite{Hassan:2020hrj}.

Using the average of 
the three results from $A$-asymmetry 
together with the ``bottle" lifetime 
$\tau_n=878.4\pm 0.5$~s, (see eq. \eqref{neutron}) gives $|V_{ud}|_n=0.97383(44)$.
\footnote{\label{foot-neut} As noted in ref. \cite{Czarnecki:2018okw}, 
the combination of eq. \eqref{Vud-super} and \eqref{neutron}, provides a precise prediction of neutron lifetime, 
which is independent of $1+\Delta^V_R$. In fact, the relation
$\tau_n = 5172.3(3.2)/(1+3g_A^2) $ is obtained, which results in perfect agreement with the ``bottle" lifetime 
$\tau_n=878.4\pm 0.5$~s 
using $g_A=-1.27624(50)$. }
By combining the results from super-allowed $\beta$ decays and free neutron decay we receive
\begin{align}
\label{eq:det_C}
|V_{ud}|_C =  0.97372(26)~,
\end{align}
We also mention that the value of $|V_{ud}|$ can also be extracted from the small branching ratio ($\sim 10^{-8}$)
measured by the PIBETA experiment of  $\pi^+\rightarrow \pi^0 e^+\nu_e$ \cite{Pocanic:2003pf}, 
which gives $|V_{ud}^\pi|=0.9739(29)$ \cite{Cirigliano:2022yyo}.
Although this decay has the cleanest theoretical prediction, the experimental uncertainty is rather large. 
Another determination is obtained also from mirror decays: $|V_{ud}^\pi|=0.9739(10)$ \cite{Hayen:2020cxh}, about $3$ times less precise than super-allowed $\beta$ decays.  
\end{itemize}

\begin{figure}[t]
\centering
\includegraphics[width=0.5\textwidth]{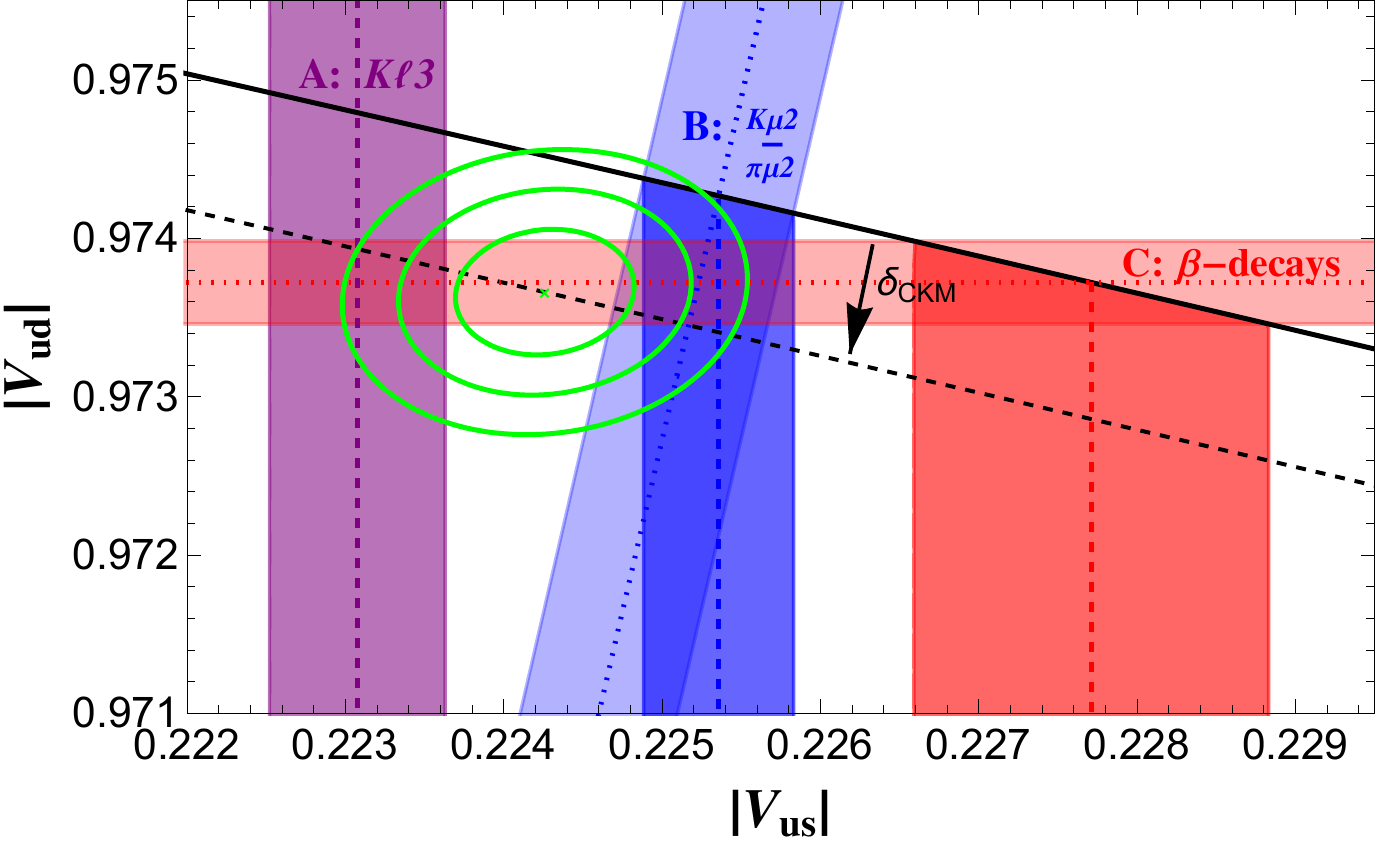} 
\caption{\label{fig:vusvud}  
Independent determinations of $|V_{us}|$ obtained from semi-leptonic $K\ell3$ decays (purple), 
$|V_{us}|/|V_{ud}|$ from leptonic $K\mu2/\pi\mu2$ decays (blue) and
$|V_{ud}|$ from super-allowed
$0^+$-$0^+$ and free neutron $\beta$ decays (red) (see text for details).
The corresponding projections on the $V_{us}$ axis are shown using the CKM first row unitarity condition \eqref{eq:uns} as depicted by the black solid line. 
A $\chi^2$ fit with two parameters $V_{us}$ and $V_{ud}$ is performed, the green curves show
$1\sigma$, $2\sigma$ and $3\sigma$ contours ($\chi^2_\text{min}+2.3,+6.18,+11.83$)
around the best-fit point.
The dashed black line corresponds to the violation of unitarity as encoded by the parameter $\delta_{\rm CKM}$.}
\end{figure}

The three determinations are not in agreement with each other within the context of the SM. The anomalies become apparent, if we translate by means of unitarity the three determinations into values of $|V_{us}|$ (or, correspondingly, of $|V_{ud}|$):
\begin{align}
\label{eq:vusabc}
& 
|V_{us}|_A = 0.22308(55)  
\, , \qquad
|V_{us}|_B = 0.22536(47) 
\, , \qquad
|V_{us}|_C = 0.2277(11)~.
\end{align}
Interestingly, there is a $3.7\sigma$ discrepancy between determination A and C. Taking a conservative average, without reducing the error, between determination A and B (obtained from kaon physics)
$|V_{us}|_{A+B} = 0.22440(51)$ still yields a $2.7\sigma$ discrepancy, which we call CAA1. It can be encoded as a deficit of the CKM first row unitarity condition by the parameter
\begin{align}
\label{eq:newundelta}
 \delta_\text{CKM} \equiv 1 - \vert V_{ud} \vert^2 - \vert V_{us} \vert^2 - \vert V_{ub} \vert^2~.
\end{align}
By performing a $\chi^2$ fit with two parameters $V_{us}$ and $V_{ud}$, using eqs. \eqref{eq:det_A}, \eqref{eq:det_B} and \eqref{eq:det_C} we obtain  $\delta_\text{CKM}\approx   1.7\times 10^{-3}$. 
Finally, CAA2 refers to the $3.1\sigma$ discrepancy between the determinations A and B obtained from kaon physics. 
We illustrate the three determinations in fig. \ref{fig:vusvud}.

\section{Vector-like quark doublet}
\label{sec:doublet}

\subsection{Model}
\label{sec:model}
In addition to the three SM chiral families of fermions, 
additional vector-like generations can exist 
with the left- and right-handed components 
in the same representations of the SM.  
Vector-like fermions are a motivated extension of the SM particle spectrum. They appear in models of grand unification \cite{Gursey:1975ki,Achiman:1978vg, Berezhiani:1989bd,Barbieri:1994kw,Berezhiani:1995dt}, play fundamental roles in models with inter-family symmetries which explain the origin of fermion mass hierarchies and mixings \cite{Berezhiani:1983hm,Dimopoulos:1983rz,Berezhiani:1985in,Berezhiani:2000cg},
 or solve the strong CP problem in models with the axion \cite{Kim:1979if,Berezhiani:1989fp} or without it 
 (Nelson-Barr type) \cite{Nelson:1983zb,Barr:1984qx,Babu:1989rb,Berezhiani:1990vp,Berezhiani:1992pq}. 
 
In the following we consider the inclusion of a vector-like weak-doublet of quarks $Q_{L,R}=( T,B)_{L,R}$ 
in the same representation of 
$SU(3)\times SU(2)_L \times U(1)_Y$ as standard left-handed quarks, that is
with the SM quantum numbers $(\bold{3},\bold{2})_{1/6}$.
The Yukawa sector is augmented by the following couplings and mass terms 
\begin{align}
\label{eq:YQ}
&\mathcal{L}_{\rm Y} \supset Y_{uij} \overline{q}_{Li} \tilde{\varphi} u_{Rj}+
Y_{dij}\overline{q}_{Li} \varphi d_{Rj}+   h_{ui}  \, \overline{Q}_{L} \tilde\varphi u_{Ri}  +  h_{di} \, \overline{Q}_{L}  \varphi  d_{Ri} + M_Q \overline{Q}_L Q_R
 ~ + ~ {\rm h.c.}~,
\end{align} 
where $\varphi$ is the Higgs doublet and $i,j=1,2,3$ are the family indexes. 
We have defined the quark basis in which the mixed mass terms of the type 
$ \mu_Q  \bar{Q}_L Q_R$ are rotated away, using the fact that the four species of left-handed doublets have identical quantum numbers.\footnote{In some predictive models for fermion masses and 
mixings, the SM Yukawa terms can be forbidden by some symmetry (e.g. flavour symmetry or Peccei-Quinn symmetry) and emerge after integrating out the heavy states as e.g. in refs.
\cite{Berezhiani:1992pj,Koide:1995pb,Koide:1999mx,Berezhiani:2005tp} (so called `universal' seesaw mechanism \cite{Rajpoot:1987ji,Davidson:1987mh,Berezhiani:1991ds}).  
In the context of supersymmetric models with flavour symmetry this mechanism can 
give a natural realization of the minimal flavour violation scenario \cite{Berezhiani:1996ii,Anselm:1996jm,Berezhiani:2001mh}.}
After substituting the Higgs VEV $\langle \phi \rangle = v_w$, the mass matrices of up type and down type quarks read
 \begin{align}\label{eq:mass}
& {\cal M}_u = \left(\begin{array}{ccc}
Y_u v_w &  0 \\
h_{u} v_w &  M_Q  
\end{array}\right)~,   \quad \quad 
{\cal M}_d = \left(\begin{array}{ccc}
Y_d v_w &  0 \\
h_{d} v_w &  M_Q  
\end{array}\right)~,
\end{align} 
where $Y_{d(u)}$ are the $3\times 3$ SM Yukawa matrices and $h_u$ and $h_d$ are row vectors $h_d = (h_{d1}, h_{d2},h_{d3})$, $h_u = (h_{u1}, h_{u2},h_{u3})$.  

The mass matrices can be diagoanlized via bi-unitary transformations 
$U_{uL}^\dagger {\cal M}_u U_{uR}=\text{diag}(y_uv_w,y_cv_w,y_tv_w,M_{T'})$ and  
$U_{dL}^\dagger {\cal M}_{d}  U_{dR}=\text{diag}(y_dv_w,y_sv_w,y_bv_w,M_{B'})$. 
The initial states are related to mass eigenstates as
\begin{align} \label{eq:mass_rot} 
& \left(\begin{array}{c} d_1 \\ d_2 \\ d_3 \\ B \end{array}\right)_{L,R}
= U_{d L,R}\left(\begin{array}{c}
d \\ s \\ b \\ B' \end{array}\right)_{L,R} \, , && 
\left(\begin{array}{c} u_1 \\ u_2 \\ u_3 \\ T \end{array}\right)_{L,R}
= U_{u L,R}\left(\begin{array}{c}
u \\ c \\ t \\ T' \end{array}\right)_{L,R}~.
\end{align}
The unitary matrices $U_{d L,R}$ can be found using the relations 
$U_{dL}^\dagger {\cal M}_{d}{\cal M}_{d}^\dag  U_{dL}=U_{dR}^\dagger {\cal M}_{d}^\dag{\cal M}_{d}  U_{dR}={\cal M}_{d,\text{diag}}^2 $ and similarly for the up sector.

As regards the left-handed rotations $U_{d(u) L}$, 
the extra elements describe the mixing of SM quarks with the vector-like doublet.
These mixings can be parameterized by sines of very small angles $s_{Li}\approx y_{i} |h_{i}| v^2_w/M^2_Q$,
proportional to the SM Yukawa couplings $y_i$ 
and suppressed by the ratio $v^2_w/M^2_Q$. 
By rotating the first three generations, we can choose the basis in which the Yukawa submatrix of up quarks $\hat{U}_{uL}^\dag Y_u\hat{U}_{uR}$ is diagonal.
Then,
the unitary matrices $U_{d(u) L}$ can be parameterized as
\begin{align}
&
U_{dL} 
=\! \left(\! \begin{array}{cccc}
U_{L1d} & U_{L1s} & U_{L1b} & U_{L1B'} \\
U_{L2d} & U_{L2s} & U_{L2b} & U_{L2B'} \\
U_{L3d} & U_{L3s} & U_{L3b} & U_{L3B'} \\
U_{LBd} & U_{LBs} & U_{LBb} & U_{LBB'} 
\end{array}\! \right)  \!
\approx \!
\left( \begin{array}{c@{\hspace{1\tabcolsep}}c@{\hspace{1\tabcolsep}}c@{\hspace{1\tabcolsep}}c}
&  &  & 0 \\
 & \hat{U}_{dL} &  & 0 \\
 & &  & 0 \\
0 & 0 & 0 & 1
\end{array} \right),
\quad
U_{uL} 
\approx  
\left( \begin{array}{c@{\hspace{2.5\tabcolsep}}c@{\hspace{1\tabcolsep}}c@{\hspace{0.5\tabcolsep}}c}
1  & 0 & 0    &0 \\
0 & 1 & 0    & 0 \\
0 & 0 &  1 & \frac{y_t h_t^* v_W^2}{M_Q^2} \\
0  & 0 & -\frac{y_t h_t v_W^2}{M_Q^2} & 1
\end{array} \! \right)~,
\label{eq:UdL}
\end{align}
where we introduced the Yukawa couplings 
$(h_u,h_c,h_t)=h_u \hat{U}_{uR}$ in this weak basis.

As regards the right-handed sector, 
we can rotate the first three generations and choose the basis in which
$\hat{U}_{dR}^\dag Y_{d}^\dag{Y}_{d}\hat{U}_{dR}$ is diagonal.
Then, for couplings 
$\lesssim \mathcal{O}(1)$,
the mixings of SM quarks with the vector-like doublet 
are determined by the small parameter $v_w/M_Q$, and we can
write $U_{dR}$ as
\begin{align}
&U_{dR}
= \left(\begin{array}{cccc}
U_{R\, 1d} & U_{R\, 1s} & U_{R\, 1b} & U_{R\, 1B'} \\
U_{R\, 2d} & U_{R\, 2s} & U_{R\, 2b} & U_{R\, 2B'} \\
U_{R\, 3d} & U_{R\, 3s} & U_{R\, 3b} & U_{R\, 3B'} \\
U_{R\, Bd} & U_{R\, Bs} & U_{R\, Bb} & U_{R\, BB'} 
\end{array}\right)
\approx
\left(\begin{array}{cccc}
1  & 0 & 0    & \frac{h^{*}_{d} v_w}{M_Q} \\
0 & 1 & 0    &  \frac{h^{*}_{s} v_w}{M_Q} \\
0& 0 & 1     &  \frac{h^{*}_{b} v_w}{M_Q} \\
-\frac{h_{d} v_w}{M_Q} & -\frac{h_{s} v_w}{M_Q} & -\frac{h_{b} v_w}{M_Q} & 1
\end{array}\right) + \mathcal{O}\Big(\frac{v_w^2}{M_Q^2}\Big)~,
\label{eq:vrd}
\end{align}
where we introduced 
$(h_d,h_s,h_b)=h_d \hat{U}_{dR}$.
Similar expressions hold for the rotations of up quarks $U_{uR}$.
As for the mass eigenvalues, we have
\begin{align}
    &  \frac{M_{T'}^2}{M_Q^2}=  1+(h_t^2+h_c^2+h_u^2) \, \frac{v_w^2}{M_Q^2} \, + \mathcal{O}\Big(\frac{ v_w^4}{M_Q^4}\Big)~,
    \label{eq:Msplit}
\end{align}
and similarly for the down sector.

Next, we discuss the modifications to the gauge interactions induced by the presence of the vector-like quark. The charged-current Lagrangian can be written as
\begin{align}
\mathcal{L}_{cc}& \supset 
- \frac{g}{\sqrt{2}} W_\mu^+  \Big[ \left(\begin{array}{cccc} 
\bar{u}_L & \bar{c}_L & \bar{t}_L & \bar{T}_L'\end{array}\right)
\gamma^\mu V_L
\left(\begin{array}{c} d_L \\ s_L \\ b_L \\ B_L' \end{array}\right) + \left(\begin{array}{cccc} 
\bar{u}_R & \bar{c}_R & \bar{t}_R & \bar{T}_R'\end{array}\right)
\gamma^\mu  
V_R\left(\begin{array}{c} d_R \\ s_R \\ b_R \\ B_R' \end{array}\right) \Big]
+\text{h.c.} 
\label{eq:Lagr_W}
\end{align}
where the right-handed current originates from the term $-\frac{g}{\sqrt{2}} W_\mu^+ \bar{T}_R \gamma^{\mu} B_R$ after performing the rotation to the mass eigenbasis (see eq. \eqref{eq:mass_rot}).

Since the four species are $SU(2)_L$ doublets, the mixing matrix for the left-handed sector is a unitary matrix. From eq. \eqref{eq:UdL} we have 
\begin{align}
&V_L=U_{uL}^{\dag}U_{dL}=\left(\begin{array}{cccc}
V_{Lud} & V_{Lus} & V_{Lub} & V_{LuB'}  \\
V_{Lcd} & V_{Lcs} & V_{Lcb} & V_{LcB'}  \\
V_{Ltd} & V_{Lts} & V_{ Ltb} & V_{LtB'} \\
V_{LT'd} & V_{LT's} & V_{LT'b} & V_{LT'B'}
\end{array}\right)
\approx \nonumber \\
&\approx
\left(\begin{array}{cccc}
V_{Lud} & V_{Lus} & V_{Lub} & 0  \\
V_{Lcd} & V_{Lcs} & V_{Lcb} &  0  \\
V_{Ltd} & V_{Lts} & V_{ Ltb} & -y_th_t^*\frac{v_w^2}{M_Q^2}+y_bh_b^*\frac{v_w^2}{M_Q^2}  \\
V_{Ltd}y_th_t\frac{v_w^2}{M_q^2} & V_{Lts} y_th_t\frac{v_w^2}{M_Q^2} &  
y_th_t\frac{v_w^2}{M_Q^2}V_{Ltb}-y_bh_b\frac{v_w^2}{M_Q^2}  & 1
\end{array}\right)~.
\label{eq:vckmL}
\end{align}
where the upper left $3 \times 3$ submatrix is given by $\hat{U}_{dL}$ and corresponds to the CKM matrix in the limit of decoupled new physics.
Then, because of the suppression of the extra mixings, it still holds for the first row the unitarity relation  
\begin{align}
  &   |V_{L ud}|^2+|V_{L us}|^2+|V_{L ub}|^2= 1 
  \label{unL}
\end{align}
where also $|V_{L ub}|$ is indeed irrelevant.

Since $Q_{R}$ is a $SU(2)_L$ doublet mixing with the right-handed singlet quarks,
a charged current coupling with the $W$ boson is generated also in the right-handed sector (see eq. \eqref{eq:Lagr_W})
 with a non-unitary
mixing matrix given by
\begin{align}
V_R&=V_{uR}^{\dag}\text{diag}(0,0,0,1)V_{dR} 
 =
\left(\begin{array}{cccc}
V_{R ud} & V_{R us} & V_{R ub} & V_{R uB'}  \\
V_{R cd} & V_{R cs} & V_{R cb} & V_{R cB'}  \\
V_{R td} & V_{R ts} & V_{R tb} & V_{R tB'} \\
V_{R T'd} & V_{R T's} & V_{R T'b} & V_{R T'B'}
\end{array}\right) \approx  \notag \\ &\approx \left(\begin{array}{cccc}
h_{u}^\ast {h}_{d} \frac{v^2_w}{M^2_q}  & h_{u}^\ast {h}_{s}\frac{v^2_w}{M^2_q}   &  h_{u}^\ast {h}_{b}  \frac{v^2_w}{M^2_q} 
& -h_{u}^\ast \frac{v_w}{M_q}   \\
h_{c}^\ast {h}_{d} \frac{v^2_w}{M^2_q}  & h_{c}^\ast {h}_{s} \frac{v^2_w}{M^2_q}   & h_{c}^\ast {h}_{b}  \frac{v^2_w}{M^2_q} 
& - h_{c}^\ast  \frac{v_w}{M_q}  \\
h_{t}^\ast {h}_{d}\frac{v^2_w}{M^2_q}  & h_{t}^\ast {h}_{s} \frac{v^2_w}{M^2_q} &  h_{t}^\ast {h}_{b}\frac{v^2_w}{M^2_q} 
 & - h_{t}^\ast   \frac{v_w}{M_q}      \\
-{h}_{d}   \frac{v_w}{M_q} & -{h}_{s}    \frac{v_w}{M_q}  &  -{h}_{b}     \frac{v_w}{M_q} &  1
\end{array}\right)~.
\label{eq:vckmR}
\end{align}
As a consequence, the weak interaction of SM quarks with the $W$-boson loses its pure $V-A$ character. 
In fact, 
by denoting as $\hat{V}_L$ and $\hat{V}_R$ 
the $3\times 3$ submatrices of $V_L$ and $V_R$ describing the mixing between SM quarks, 
from eq. \eqref{eq:Lagr_W} the couplings change as
  \begin{align}\label{cc3} 
\mathcal{L}_{cc}& \supset  -\frac{g}{2\sqrt2}\, W^+_\mu \, \overline{(u  ~~ c   ~~ t ) } \,  
 \big[ \gamma^\mu (\hat{V}_L+ \hat{V}_R) - \gamma^\mu \gamma^5 (\hat{V}_L-\hat{V}_R ) \big]
 \, \left(\begin{array}{c} d  \\ s  \\ b   \end{array}\right)~.
\end{align}  

Additionally, the mixing of SM quarks with the vector-like quarks induces 
couplings with Higgs and $Z$-boson which are flavour-non-diagonal in the mass basis and originate flavour-changing phenomena.
In particular,
left-handed couplings remain diagonal at tree level as in the SM,
while in the right-handed sector additional couplings proportional to weak isospin appear
\begin{align}
\label{eq:Lagr_Z}
\mathcal{L}_\text{nc} \supset &-\frac{g}{\cos\theta_W}Z^\mu\left(\begin{array}{cccc}
\overline{u}_R & \overline{c}_R & \overline{t}_R & \overline{T'}_R 
\end{array}\right)\gamma^\mu \left(\frac{1}{2}K_{uR}-\frac{2}{3}\sin^2\theta_w\mathbf{1}\right) \left(\begin{array}{c}
u_R \\ c_R \\ t_R \\ T'_R
\end{array} \right)- \nonumber  \\
&-\frac{g}{\cos\theta_W}Z^\mu\left(\begin{array}{cccc}
\overline{d}_R & \overline{s}_R & \overline{b}_R & \overline{B'}_R 
\end{array}\right)\gamma^\mu \left(-\frac{1}{2}K_{dR}+\frac{1}{3}\sin^2\theta_w\mathbf{1}\right) \left(\begin{array}{c}
d_R \\ s_R \\ b_R \\ B'_R
\end{array} \right)~,
\end{align}
where 
\begin{align}
&K_{uR}= U_{uR}^{\dag}\text{diag}(0,0,0,1)U_{uR} \, , \qquad
K_{dR}= U_{dR}^{\dag}\text{diag}(0,0,0,1)U_{dR} \, .
\label{eq:vnc}
\end{align}
As far as the interactions with the radial Higgs $H$ are concerned, 
the mass matrices are not proportional to the Yukawa matrices and flavour non-diagonal couplings are also generated.
 The relevant Lagrangian reads
\begin{align}
\label{eq:Lagr_H}
\mathcal{L}_H \supset \frac{1}{\sqrt{2}} H \left(\begin{array}{cccc}
\overline{d}_L & \overline{s}_L & \overline{b}_L & \overline{B'}_L 
\end{array}\right)U_{dL}^\dag 
\left(\begin{array}{cc}
 Y_{d} &  0 \\
h_{d}  & 0
\end{array}\right)
U_{dR}\left(\begin{array}{c}
d_R \\ s_R \\ b_R \\ B'_R
\end{array} \right)~,
\end{align}
and similarly for up-type quarks.

\subsection{Observables}
\label{sec:pheno}

\subsubsection{Cabibbo angle anomalies}
\label{sec:CAA}

\begin{figure}[t]
\centering
\includegraphics[width=0.6\textwidth]{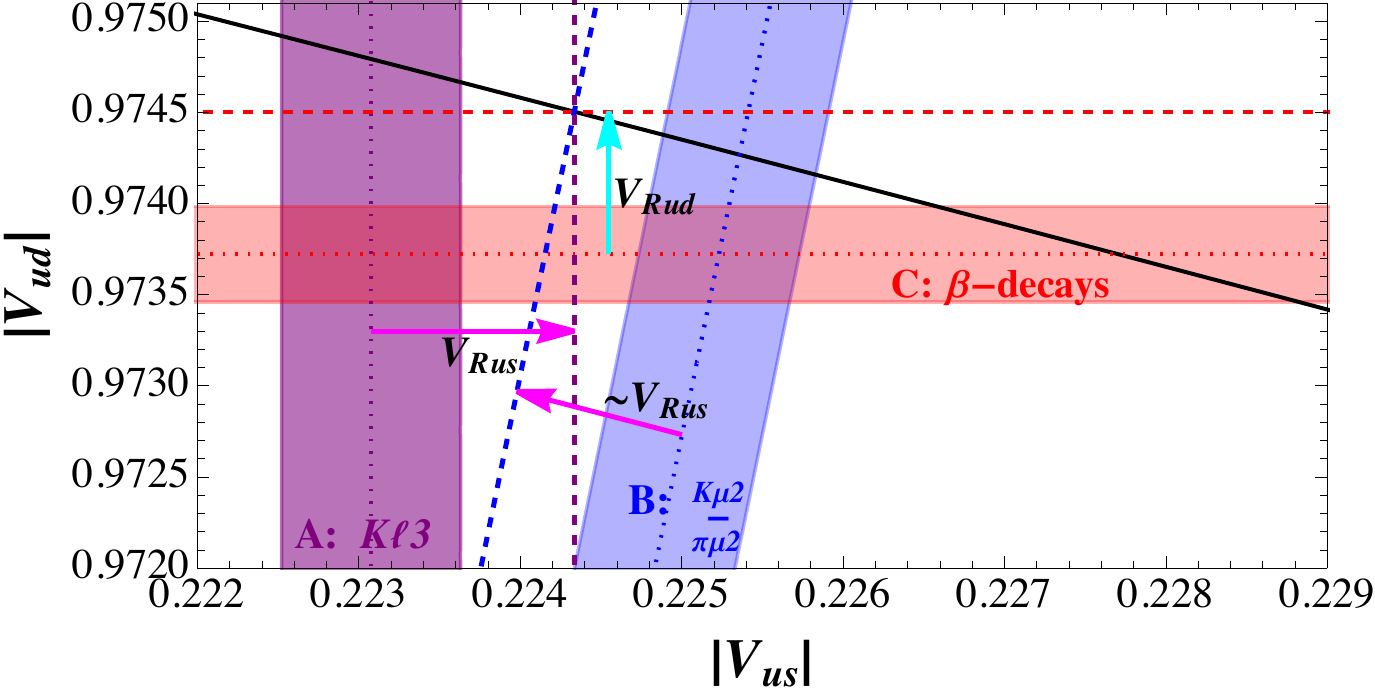} 
\caption{\label{vrudvrus} 
Effect of the mixing of the vector-like quark doublet with the SM quarks on the three independent determinations of
$|V_{us}|$ and $|V_{ud}|$.
Semi-leptonic $K\ell3$ decays (purple) provide the vector coupling $|V_{Lus}+V_{Rus}|$,
$\beta$ decays (red) the vector coupling $|V_{Lud}+V_{Rud}|$ while
leptonic decays $K\rightarrow \mu\nu/\pi\rightarrow \mu\nu$ (blue) the axial-vector couplings
$|V_{Lus}-V_{Rus}|/|V_{Lud}-V_{Rud}|$.
The black solid line depicts the 
CKM first row unitarity condition in eq. \eqref{eq:uns}. The dashed lines represent the elements $V_{Lud}$, $V_{Lus}$, and $V_{Lus}/V_{Lud}$, while the arrows the solutions to CAA1 (cyan) and CAA2 (magenta) (see eq. \eqref{eq:sol}).}
\end{figure}

The value of $V_{ud}$ obtained from $\beta$ decays
is determined by the vector coupling  $G_V = G_F \vert V_{ud} \vert$.
Also semi-leptonic kaon decays $K\ell 3$ determine the weak vector coupling, while leptonic decays $K\mu2$ and $\pi\mu2$ depend on the axial-vector current. 
As stated in the previous section, 
in the scenario with vector-like doublet vector and axial-vector couplings are not equal as in the SM.
Therefore, the three determinations correspond to different couplings \cite{Belfatto:2021jhf}.
From eqs. \eqref{eq:Lagr_W}, \eqref{eq:vckmR} we have 
\begin{align}
& |V_{us}|_A =
|V_{L us}+V_{R us}| = 0.22308(55)~,  \nonumber
\\
& |V_{us}|/|V_{ud}|_B = 
\frac{|V_{L us}-V_{R us}|}{|V_{L ud}-V_{R ud}|}=0.23131(51)~,  \label{eq:ABC} \\
&|V_{ud}|_C =
|V_{L ud}+V_{R ud}| =   0.97372(26)~. \nonumber
\end{align}
This system can be solved with real parameters
\begin{align}
&V_{R ud}= U_{R\, Tu}^*U_{R\, Bd} \approx h_{u}^\ast {h}_{d} \frac{v^2_w}{M^2_Q} =-0.78(27) \times 10^{-3} ~,  \notag \\ &V_{R us}=U_{R\, Tu}^*U_{R\, Bs} \approx h_{u}^\ast {h}_{s}\frac{v^2_w}{M^2_Q} = - 1.26(38) \times 10^{-3}~,
\label{eq:sol}
\end{align}
and $ V_{L ud} = 0.97450(8)$, $V_{L us}=0.22434(36)$, using the unitarity of the $V_L$ matrix.

Consequently, the mixing $V_{Rud}$ in the right-handed current
can explain the apparent deficit in CKM unitarity
when confronting determination from $\beta$ decays 
with the other determinations from kaon decays,
while $V_{Rus}$ 
would explain the gap between the determinations
from semi-leptonic kaon decays $K\ell 3$ and leptonic decays $K\mu2/\pi\mu 2$. This is the special property of the contributions generated by the vector-like quark doublet which, by generating right-handed currents,
induces a difference in vector and axial-vector couplings for each transition. 
From eq. \eqref{eq:ABC}, it also results that in order to have $h_i\lesssim 1$, 
it should be $M_Q\lesssim 6$~TeV for the mass of the vector-like species.

\subsubsection{$W$-boson mass}
\label{sec:CAA}

The recent measurement of the mass of the $W$ boson published by the CDF Collaboration~\cite{CDF:2022hxs} 
 $m_{W,{\rm CDF II}} = 80.4335 \pm 0.0094$~GeV exhibits a discrepancy of around $6.6\sigma$ from the SM expectation $m_{W,\text{SM}} = 80.360 \pm 0.006 $~GeV~\cite{Workman:2022ynf}. There is also a $3.7\sigma$ discrepancy with the average 
 of the other measurements $m_{W,\text{old}} = 80.377 \pm 0.012 $~GeV~\cite{Workman:2022ynf}.

 In ref. \cite{deBlas:2022hdk} an average of all measurements including the CDF-II result has been computed
 \begin{align}
 & m_{W,\text{exp}}=80.413 \pm 0.015~.
 \label{eq:mwav}
 \end{align}
 
The vector-like quark doublet 
induces radiative corrections to gauge bosons propagators (oblique corrections) via one loop diagrams, which can be parameterized by the electroweak oblique parameters $T$, $S$ and $U$ \cite{Peskin:1991sw}.
The shift of the $W$ mass
in terms of the oblique parameters reads \cite{Peskin:1991sw}
\begin{align}
\label{eq:deltamW}
\delta m^2_W=m_{W}^2-m_{W,\text{SM}}^2=c^2m_Z^2\,\alpha\,\Big[\frac{ c^2}{c^2-s^2} \,T-\frac{1}{2(c^2-s^2)}\, S+
\frac{1}{4s^2} \, U \Big]~,
\end{align}
where  $m_{Z}$ is the $Z$-boson mass, $m_{Z}=91.1882(20)$~GeV, $s=\sin\theta_W$, $c=\cos\theta_W$ with $\theta_W$ the Weinberg angle,
$s^2=0.23122(4)$, and $\alpha$ is the fine structure constant $\alpha(m_Z)=1/127.951(9)$.
We report in appendix \ref{app:oblique} the general expressions of $T$, $S$ and $U$ in presence of vector-like quarks as derived in ref. \cite{Lavoura:1992np}.
In the scenario with the vector-like doublet, 
taking into account the matrices $V_L$, $V_R$ and $K_{d(u)R}$ 
in eqs. \eqref{eq:vckmL}, \eqref{eq:vckmR}, and \eqref{eq:vnc},
and the mass splitting given in eq. \eqref{eq:Msplit},
after subtracting the SM effect of the top quark, 
 the contribution to the $T$ parameter results in
\begin{align}
\label{eq:T}
 T &\approx  
\frac{3}{16\pi^2}\frac{2\pi \, v_w^2}{s^2c^2\, m_Z^2} 
\bigg[ y_t^2|h_t|^2\frac{v_w^2}{M_Q^2} \Bigg( -3+ 2\ln\frac{M^2_Q}{m_t^2}\Bigg)+ \nonumber \\&  +
\frac{2}{3}\frac{v_w^2}{M_Q^2} \left(  \sum_{\alpha=u,c,t} |h_\alpha|^2 - \sum_{\beta =d,s,b} |h_\beta |^2 \right)^2 +\mathcal{O}\left(\frac{v_w^4}{M_Q^4} \right)
\bigg]~, 
\end{align}
For the $S$ and $U$ parameters we have
\begin{align}
\label{eq:S}
 S  &\approx \frac{3}{18\pi}  \frac{v_w^2}{M_Q^2}  \Bigg[  
  \sum_{\alpha =u,c,t} |h_\alpha|^2 \Bigg(\! -10+4\ln\frac{M^2_Q}{m_\alpha^2}\Bigg) 
  +\sum_{\beta =d,s,b} |h_\beta |^2 \left(\! -6+2\ln\frac{M^2_Q}{m_\beta^2}\right) 
\Bigg] +\, \mathcal{O}\left(\frac{v_w^4}{M_Q^4} \right)~,
\\
\label{eq:U}
 U  &\approx \frac{1}{2\pi}  \frac{v_w^2}{M_Q^2}  \Big[  
( |h_u|^2+|h_c|^2+|h_t|^2+ |h_d|^2+|h_s|^2+|h_b|^2 ) 
 +4.2\,\text{Re}(V_{Lud}h_uh_d^*)
\Big] +\, \mathcal{O}\left(\frac{v_w^4}{M_Q^4} \right).
\end{align}
The contribution of the parameter $U$ is negligible in this scenario.
Since the coupling with the top $h_t$ is larger than 
the couplings to lighter quarks, the $T$ parameter produces the main contribution to $\delta m^2_W$.
Then, the shift in the $W$ mass is mostly originated by the weak isospin-breaking effect
\begin{align}
\frac{\delta m^2_W}{m_{W,\text{SM}}^2}&\approx \frac{ c^2}{c^2-s^2} 
\frac{3}{16\pi^2} \, \frac{v_w^2}{M_Q^2}\,
\bigg[ y_t^2|h_t|^2  \Bigg( -3+ 2\ln\frac{M^2_Q}{m_t^2}\Bigg) + \nonumber \\
& +\frac{2}{3} \left(|h_d|+|h_s|^2+|h_b|^2-|h_u|^2-|h_c|^2-|h_t|^2 \right)^2 
\bigg]~.
\end{align}
In the scenario of one vector-like doublet with mass $M_Q=2$~TeV coupling only to the top,
the $m_W$ value of eq. \eqref{eq:mwav} would be explained with a Yukawa coupling of $h_t\approx 1.0\pm 0.1$, corresponding to a mass splitting of $~7$~GeV. Assuming the CDF II result, we would get $h_t\approx 1.1  \pm 0.1$. 

\subsection{Low-energy constraints}
\label{sec:constraints}

In the following, we mention the most stringent constraints from flavour-changing phenomena and then present the ones from flavour-conserving observables and processes involving the third family. In Table \ref{tbl:constraints} we summarize the most important constraints on the model parameters.

\subsubsection{Flavour-changing neutral currents}
\label{sec:FCNCs}

Flavour-changing neutral currents (FCNCs) are rare processes within the SM because they appear only at loop level and receive additional suppression due to the Glashow-Iliopoulos-Maiani mechanism~\cite{Glashow:1970gm,Glashow:1976nt,Paschos:1976ay}. The mixing of SM quarks with the vector-like quarks, which generates flavour non-diagonal couplings with the Higgs and the $Z$ boson, is thus strongly constrained by FCNCs (e.g. see also refs. \cite{Lavoura:1992qd,Ishiwata:2015cga}). 

The most stringent constraints come from kaons ($K^0$-$\bar{K}^0$ mixing, 
$K^+\rightarrow \pi^+\nu\bar{\nu}$, $K_\text{L}\rightarrow \mu^+\mu^-$, etc.) and we report the leading ones in appendix \ref{app:fcnc} in more detail. Considered together with flavour-diagonal constraints, it was shown in ref. \cite{Belfatto:2021jhf}  that one vector-like doublet cannot actually accommodate both Cabibbo angle anomalies simultaneously. In particular, 
these processes impose a bound on the product of the mixing elements 
$U_{RBd}$, $U_{RBs}$ in eq. \eqref{eq:vrd}. 
Depending on the phase $\text{Arg}(U_{RBs}^*U_{RBd})$, 
assuming $M_Q=2$~TeV for the mass of the vector-like doublet we receive the estimate
\begin{align}
&  |U_{RBs}^{*}U_{RBd}|\approx  |h_{s}^{*}h_{d}|\frac{v^2_w}{M^2_Q} \lesssim (0.6~\mbox{---}~5.2) \times 10^{-6}~,
\label{eq:vdsdoub2}
\end{align}
and this constraint becomes even more stringent for larger values of the mass of the extra doublet.
On the other hand, according to eq. \eqref{eq:sol} a solution to the anomalies would require
\begin{align}
& |U_{RTu}|^2U_{RBs}^{*}U_{RBd} \approx 1.0 \times 10^{-6}~.
\end{align}
The less stringent upper bound in eq. \eqref{eq:vdsdoub2} applied to this relation leads to $|U_{RTu}|  \approx |h_u| v_w/M_Q \gtrsim 0.3$. Such a large mixing with the heavy species contradicts data on $Z$ decay rate into hadrons, which imply $|U_{RTu}| < 0.08$, as we will see in section \ref{sec:flavour-diagonal}.
Moreover, such a mixing also implies a large Yukawa coupling $h_u\approx 1.8 \, (M_Q/1\,\text{TeV})$,
at the verge of loss of perturbativity. 
If the vector-like doublet couples predominantly to either the down or the strange quark, while the other coupling is suppressed,
flavour-changing effects can be avoided.
Following this observation, in section \ref{sec:results}, we are addressing only one Cabibbo Angle anomaly at a time assuming that some other new physics would be necessary to explain the other. 

Mixing in the neutral $B^0_{d(s)}$-mesons system and flavour-changing $B_{d(s)}$-decays impose constraints on the mixing $|U_{RBd(s)}^*U_{RBb}|$. We receive the approximate bounds
$|U_{RBd}^*U_{RBb}|\approx |h_{d}^{*}h_{b}|v^2_w/M_Q^2< (0.4 \mbox{---} 1.7) \times 10^{-4} $ and
$|U_{RBs}^*U_{RBb}|\approx |h_{s}^{*}h_{b}|\frac{v^2_w}{M_Q^2}< 8 \times 10^{-7}  \mbox{---} 6.4 \times 10^{-4}$. Regarding the up-quarks sector, we estimate from the neutral $D$-mesons system that $|U^*_{RTu}U_{RTc}|\approx |h_{u}^*h_{c}|v_w^2/M_Q^2< 1.0 \times 10^{-4} $ for the mass of the vector-like doublet $M_Q=2$~TeV. We will assume that the couplings to charm and bottom are small enough to respect these limits.

\subsubsection{Top decays}
\label{eq:top_dec}

Since couplings of the vector-like doublet with both top quark and light quarks are different from zero, 
flavour-changing top decays $t\rightarrow Hu$, $t\rightarrow Zu$ are induced. 
The experimental limits on these decays are \cite{ATLAS:2021stq,ATLAS:2022gzn}
\begin{align}
&\text{Br}(t\rightarrow uZ)_{\rm exp}<6.6 \times 10^{-5} \, , \qquad \text{Br}(t\rightarrow cZ)_{\rm exp}<1.2 \times 10^{-4}~, \\
&\text{Br}(t\rightarrow uH)_{\rm exp}<6.9 \times 10^{-4}\, , \qquad \text{Br}(t\rightarrow cH)_{\rm exp}<9.4 \times 10^{-4}~.
\end{align}
The SM predictions for these flavour-changing decays are well below experimental bounds, 
$\text{Br}(t\rightarrow uZ)_\text{SM}\sim 10^{-16}$, $\text{Br}(t\rightarrow uH)_\text{SM}\sim 10^{-17}$, $\text{Br}(t\rightarrow cZ)_\text{SM}\sim 10^{-14}$ and $\text{Br}(t\rightarrow cH)_\text{SM}\sim 10^{-15}$
\cite{Aguilar-Saavedra:2004mfd}.
In our scenario the Lagrangian includes the interaction terms
\begin{align}
\mathcal{L}_{\rm top} \supset -\frac{g}{2\cos\theta_W} U_{RTu}^*U_{RTt} Z^\mu \bar{u}_R\gamma_\mu t_R - \frac{1}{\sqrt{2}} \,U_{LTt}\,h_u^* H \bar{u}_Rt_L +\text{h.c.}~,
\end{align}
(and similarly for charm, however we neglect the couplings with second generation in our case).
The predicted branching ratios are
\begin{align}
&\text{Br}(t\rightarrow uZ)_\text{NP}\approx \frac{1}{2|V_{tb}|^2} \big|U_{RTu}^*U_{RTt}\big|^2 
\left(1-\frac{m_Z^2}{m_t^2}  \right)^2 \left(1+2\frac{m_Z^2}{m_t^2}  \right)
\left(1-\frac{m_W^2}{m_t^2}  \right)^{-2} \left(1+2\frac{m_W^2}{m_t^2}  \right)^{-1}~, \\
&\text{Br}(t\rightarrow uH)_\text{NP}\approx \frac{1}{\Gamma_t}\frac{1}{64\pi}\big|U_{LTt}\,h_u^*\big|^2 
m_t\left(1-\frac{m_H^2}{m_t^2}  \right)^2~, 
\end{align}
using the rate $\Gamma_t \approx \Gamma (t\rightarrow bW^+)=\frac{G_F}{8\pi\sqrt{2}}|V_{tb}|^2m^3_t\left(1-\frac{m_W^2}{m_t^2}  \right)^2 \left(1+2\frac{m_W^2}{m_t^2}  \right)$.
The resulting bounds are
\begin{align}
&\big|U_{RTu}^*U_{RTt}\big|\approx|h_u^*h_t| \frac{v_w^2}{M_Q^2} \lesssim 0.01~, \qquad 
\big|U_{LTt}\,h_u^*\big|\approx y_t|h_u^*h_t|\frac{v_w^2}{M_Q^2} \lesssim 0.08~.
\end{align}

We also report here the future prospects for these decay channels in High Luminosity LHC (HL-LHC) and the hadron-hadron Future Circular Collider (FCC-hh). The center of mass energy and integrated luminosity for HL-LHC are $14~\rm TeV$ and $3~\rm ab^{-1}$~\cite{Liu:2020bem,ATLAS:2016qxw} and for FCC-hh $100~\rm TeV$ and $30~\rm ab^{-1}$~\cite{Liu:2020kxt,Liu:2020bem}. For the FCC-hh we give both the limits assuming conservative $10 \%$ systematics as well as optimistic $0 \%$ in the parenthesis. 
\begin{align}
&\text{Br}(t\rightarrow uZ)_{\rm HL-LHC}< 4.08 (2.34) \times 10^{-5} \, , && \text{Br}(t\rightarrow cZ)_{\rm HL-LHC}< 6.65(3.13)  \times 10^{-5}~, \notag \\
&\text{Br}(t\rightarrow uH)_{\rm HL-LHC}<2.4 \times 10^{-4}\, , && \text{Br}(t\rightarrow cH)_{\rm HL-LHC}<2 \times 10^{-4}~,
\end{align}
\begin{align}
&\text{Br}(t\rightarrow uZ)_{\rm FCC-hh}<2.17 (0.069) \times 10^{-5} \, , && \text{Br}(t\rightarrow cZ)_{\rm FCC-hh}<3.54 (0.089) \times 10^{-5}~, \notag \\
&\text{Br}(t\rightarrow uH)_{\rm FCC-hh}<2.3(0.73) \times 10^{-5}\, , && \text{Br}(t\rightarrow cH)_{\rm FCC-hh}<3(0.96) \times 10^{-5}~.
\end{align}

\subsubsection{Flavour-conserving processes}
\label{sec:flavour-diagonal}

First, we consider the total decay rate of the $Z$ boson as well as the partial decay rate into hadrons. The experimental measurements yield \cite{ALEPH:2005ab, Workman:2022ynf}
\begin{align}
\Gamma(Z)_\text{exp}=2.4955\pm0.0023~\text{GeV}~, \qquad \Gamma(Z\rightarrow\text{hadr})_\text{exp}=1.7448\pm0.0026~\text{GeV}~,
\end{align}
while the corresponding SM predictions are $\Gamma(Z)_\text{SM}=2.4941\pm0.0009~\rm GeV$ and $\Gamma(Z\rightarrow\text{hadr})_\text{SM}=1.74097\pm0.00085~\rm GeV$~\cite{Workman:2022ynf}.

Mixing with the heavy doublet changes the prediction as  
\begin{align}
&\Gamma(Z\rightarrow\text{had})-\Gamma(Z\rightarrow\text{had})_\text{SM}=
 \Gamma(Z)-\Gamma(Z)_\text{SM}
\approx  \nonumber \\
\approx&\frac{G_FM^3_Z}{\sqrt{2}\pi}   \left[
-\frac{2}{3}\sin^2\theta_W\left(|U_{RTu}|^2+|U_{RTc}|^2\right)
-\frac{1}{3}\sin^2\theta_W \left(|U_{RBd}|^2+|U_{RBs}|^2+|U_{RBb}|^2\right)\right]~,
\label{eq:Zdo}
\end{align}
which means that the predicted decay rate is lower than the SM expectation.
At $2\sigma$ CL we obtain the limit 
\begin{align}
&
|U_{RTu}|^2+|U_{RTc}|^2+\frac{1}{2} \left(|U_{RBd}|^2+|U_{RBs}|^2+|U_{RBb}|^2\right) \lesssim 5 \times 10^{-3}~.
\end{align}

Another set of flavour-conserving constraints originate from parity violating effects at low-energy electron-hadron processes with $Z$-boson exchange. The interaction Lagrangian can be written as
\begin{align}
\mathcal{L}_{e-\rm{had}} =
\frac{G_F}{\sqrt{2}}\sum_q \left(g^{eq}_{AV}\bar{e}\gamma_\mu\gamma^5 e\bar{q}\gamma^\mu q+g^{eq}_{VA}\bar{e}\gamma_\mu e\bar{q}\gamma^\mu\gamma^5 q\right)~.
\end{align}
Measurements of 
atomic parity violation provide the determination of nuclear weak charges $Q^{Z,N}_W$ \cite{Workman:2022ynf}
\begin{align}
\label{QWZN}
& Q^{Z,N}_W=-2[Z(g^{ep}_{AV}+0.00005)+N(g^{en}_{AV}+0.00006)](1-\frac{\alpha}{2\pi}),
\end{align}
where $Z$ and $N$ are the 
numbers of protons and neutrons in the nucleus,
$g^{ep}_{AV}=2 g^{eu}_{AV}+g^{ed}_{AV}$, $g^{en}_{AV}= g^{eu}_{AV}+2g^{ed}_{AV}$,
and $\alpha$ is the fine structure constant, $\alpha^{-1}\approx 137.036$. After including higher orders corrections, the value of the neutral current parameters is
$g^{eu}_{AV,\text{SM}}=-0.1887$ and  $g^{ed}_{AV,\text{SM}}=0.3419$ \cite{Workman:2022ynf}.
The most precise measurement of atomic parity violation is in Cesium
\cite{Workman:2022ynf}
\begin{align}
& Q^{55,78}_W(Cs)_\text{exp}=-72.82\pm 0.42~,
\end{align}
corresponding to $55 g^{ep}_{AV}+78 g^{en}_{AV}=36.46\pm 0.21$, 
while the SM prediction is $Q^{55,78}_W(Cs)_\text{SM}=-73.24\pm 0.01$ 
(i.e. $55 g^{ep}_{AV,\text{SM}}+78 g^{en}_{AV,\text{SM}}=36.66 $) \cite{Workman:2022ynf}.
The new physics contribution to the weak charge of Cesium in our scenario is given by
\begin{align}
& \Delta Q_W(Cs)=Q_W^{55,78}(Cs)-Q_W^{55,78}(Cs)_\text{SM} \approx  -2\left(-94 \frac{1}{2} |U_{RTu}|^2+105.5  |U_{RBd}|^2\right)~.
\end{align}
The parity violating asymmetry in 
$e^-p\rightarrow e^-p$ elastic scattering was determined by the 
$Q_\text{weak}$ collaboration \cite{Qweak:2018tjf},
resulting in the experimental determination of the weak charge of the proton 
$Q_{W}(p)_{\text{exp}} = 0.0719\pm 0.0045$ \cite{Workman:2022ynf},
which corresponds to the coupling 
\begin{align}
&g^{ep}_{AV,\text{exp}}=- 0.0356 \pm 0.0023~.
\end{align}
In the SM, we have $Q_{W}(p)_{\text{SM}} = 0.0709\pm 0.0002$
(i.e. $g^{ep}_{AV,\text{SM}}=-0.0355 $)
in agreement with the experimental result. 
The new contribution to $ g^{ep}_{AV}$ due to the presence of the vector-like doublet is
\begin{align}
\Delta g^{ep}_{AV}=g^{ep}_{AV}-g^{ep}_{AV,\text{SM}} \approx
-|U_{RTu}|^2+\frac{1}{2}|U_{RBd}|^2~.
\end{align}

\begin{table*}
\centering
\begin{tabular}{| c | c |}
\hline
\bf{Process} & \bf{Constraint}  \\
\hline
$\Gamma(Z\rightarrow \text{hadrons})$, $\Gamma(Z)$ & 
$\left[
|U_{RTu}|^2+|U_{RTc}|^2+0.5 \sum_{q=d,s,b}|U_{RBq}|^2\right]  \lesssim 5 \times 10^{-3}$
 \\ 
$Q_W(Cs)$ & 
$-0.0022 
< (|U_{RTu}|^2-1.12|U_{RBd}|^2)< 0.0066 
$  \\ 
$Q_W(p)$ & $ \left| -|U_{RTu}|^2+\frac{1}{2}|U_{RBd}|^2 \right| <0.0045$    \\ 
\hline
$t\rightarrow uH$, $t\rightarrow cH$ &  $|U_{RTu,c}^*U_{RTt}|m_t/v_w\lesssim 0.08$     \\
$t\rightarrow uZ$, $t\rightarrow cZ$ &     
$|U_{RTu,c}^*U_{RTt}| \lesssim 0.01$  \\
\hline
$K^+\rightarrow \pi^+\nu\bar{\nu}$,  $K_\text{L}\rightarrow \mu^+\mu^-$, $K^0$-$\bar{K}^0$  & $|U_{RBs}^{*}U_{RBd}|  \lesssim 5 \times 10^{-6} $ \\
\hline
$B^0$-$\bar{B}^0$, $B^0\rightarrow\mu^+\mu^-$  &  $ |U_{RBd}^{*}U_{RBb}| < 1.8 \times 10^{-4} $ 
\\
$B^0_s$-$\bar{B}^0_s$, $B^0_s\rightarrow\mu^+\mu^-$  &  $ |U_{RBd}^{*}U_{RBb}| < 3.3 \times 10^{-4} $ 
\\
\hline
$D^0$-$\bar{D}^0$ & 
$  |U_{RTu}^{*}U_{RTc}|< 1.3 \times 10^{-4}  $ \\
\hline
\end{tabular}
\caption{\label{tbl:constraints} 
Summary of most relevant experimental bounds on the mixing of the vector-like quark doublet with the SM quarks.}
\end{table*}

\section{Discussion}
\label{sec:discussion}

In order to alleviate the flavour-changing constraints discussed in section \ref{sec:FCNCs}, in this work we consider two separate scenarios, based on whether the vector-like doublet couples predominantly to the down or the strange quark. Consequently, the two scenario aim towards an explanation of either CAA1 or CAA2 respectively,
together with the CDF-II result. In the following, we present the two scenarios, reporting in the parenthesis at the beginning of the paragraph the relevant new physics couplings.\footnote{We use real parameters for simplicity.}

\subsection{Vector-like quark doublet}
\label{sec:results}
\begin{figure}[t]
\centering
\begin{subfigure}{0.49\textwidth}
\includegraphics[width=\textwidth]{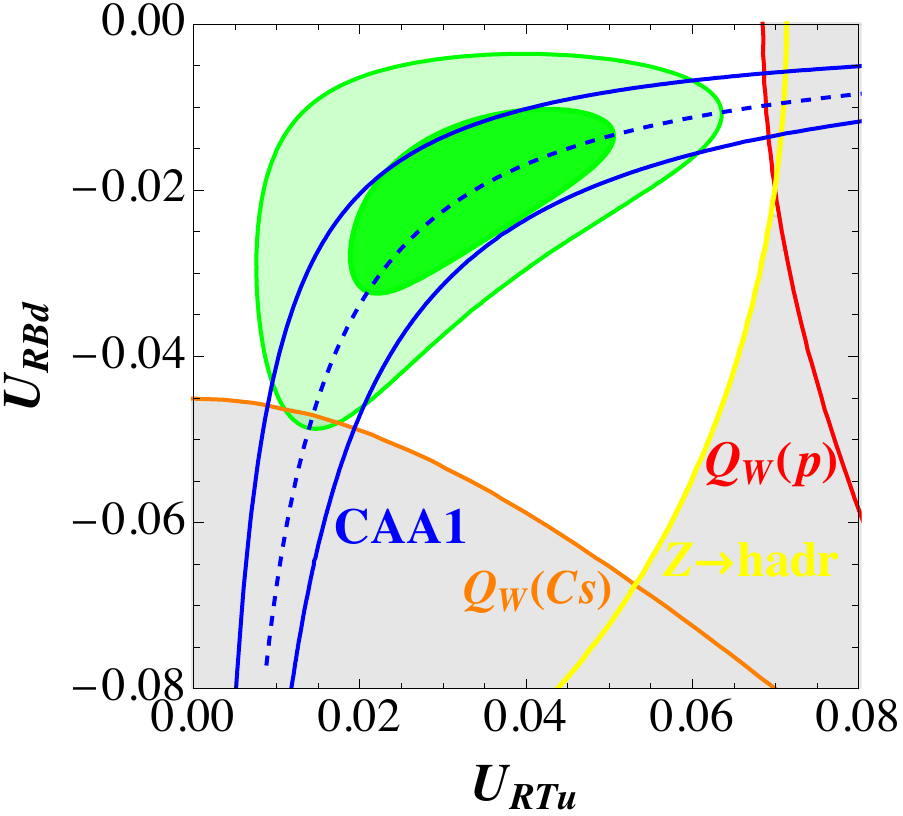}
\caption{\label{fig:fit2TeVdown1}}
 \end{subfigure}
\hfill
\begin{subfigure}{0.47\textwidth}
\includegraphics[width=\textwidth]{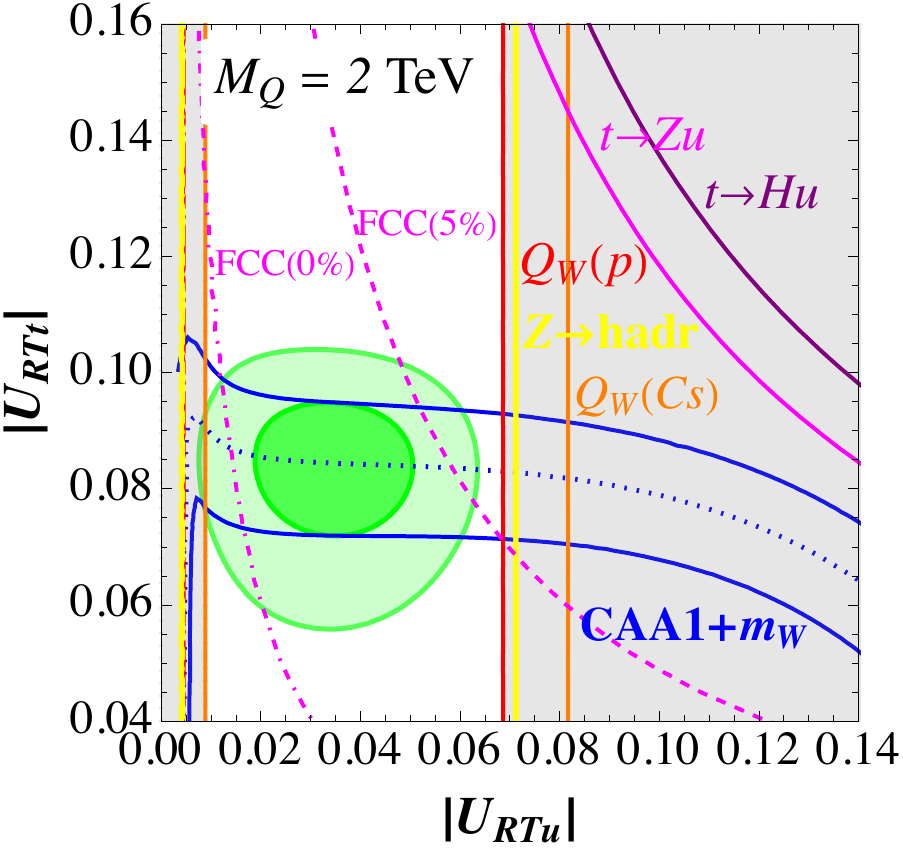}
\caption{\label{fig:fit2TeVdown2}}
 \end{subfigure}
\hfill
\begin{subfigure}{0.4\textwidth}
\includegraphics[width=\textwidth]{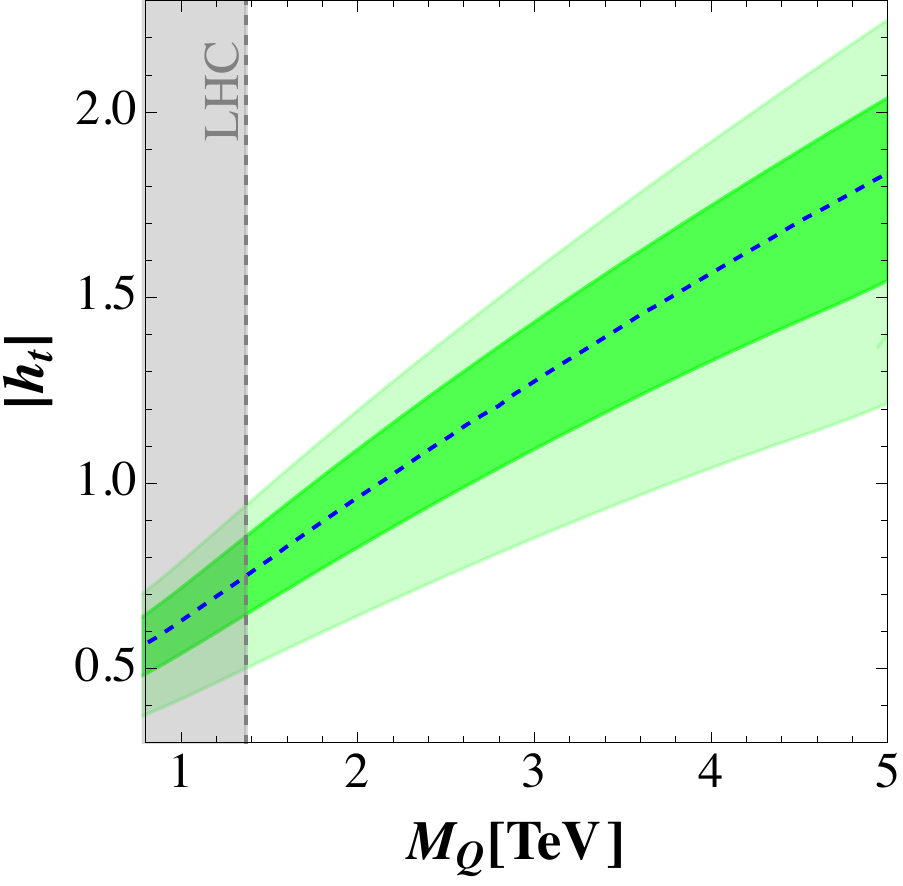}
 \caption{ \label{fig:fit2TeVdown3}}
\end{subfigure}
\caption{\label{fig:fit2TeVdown} Parameter space in the scenario with one vector-like quark doublet coupling with
up, down and top quarks (see eq. \eqref{eq:vrd}). 
$1\sigma$ and $2\sigma$  
preferred regions of 
the mixing parameters are indicated (green and lighter green),
($\chi^2_\text{min}+1$, $\chi^2_\text{min}+4$).
We also show the region excluded at $2\sigma$ by experimental bounds (grey) and
$1\sigma$ interval 
which would explain the CAA1 and the $m_W$-mass shift (blue band), without including other constraints.
In figure \ref{fig:fit2TeVdown2}, $M_Q=2\,\text{TeV}$ is assumed for the mass of the vector-like doublet 
and we indicate the experimental bounds on the left side using the conservative value $|V_{RTd}|=0.4\times 10^{-3}$ (grey region on the left side).}
\end{figure}

\paragraph{
\textbf{CAA1 + $\bm{m_W}$ ($\bm{h_u,~h_d,~h_t,~M_Q}$):}}
We perform a $\chi^2$ fit using the parameters $U_{RTu}$, $U_{RBd}$, $U_{RTt}$, $V_{Lus}$, including the Cabbibo angle determinations of eqs. \eqref{eq:det_A}, \eqref{eq:det_B}, \eqref{eq:det_C}, the $W$-mass global value of eq. \eqref{eq:mwav} and the observables in sec. \ref{sec:flavour-diagonal}  (i.e. $\Gamma(Z \to \rm had)$, $Q_W(p)$, $Q_W(Cs)$).  
In this scenario, the best fit point results in
\begin{align}
 U_{RTu}= \pm 0.035~, \quad   U_{RBd}=\mp 0.019~, \quad |U_{RTt}|=0.084~, \quad V_{Lus}=0.22452~,
\end{align}
where $U_{RTt}$ is determined for $M_Q=2$~TeV.
The SM fit assuming CKM unitarity yields in the minimum
$\chi^2_{\text{SM}}=31.4$ with  $V_{us}=0.2247$.
We obtain an improvement of $\chi^2_\text{SM}-\chi^2_\text{min}=17.2$ which is independent of the mass of the vector-like doublet.
The remaining discrepancy is due to the difference between the determinations of Cabibbo angle from $K\ell3$ and $K\mu2/\pi\mu2$ decays. 
 If this discrepancy is resolved by the presence of another vector-like doublet, we would obtain $\chi^2_\text{SM}-\chi^2_\text{min}=28.1$.

We present the results of the analysis in the $U_{RTu}$-$U_{RBd}$ and $U_{RTu}$-$U_{RTt}$
planes in figures \ref{fig:fit2TeVdown1} and \ref{fig:fit2TeVdown2}
respectively,
marginalizing over the other variables.
Limits from LHC exclude vector-like doublets coupling to the top for masses lower than
$~\sim1.37$~TeV \cite{ATLAS:2018ziw}, so figure \ref{fig:fit2TeVdown2} is given for a benchmark mass $M_Q=2$~TeV.
The plot in figure \ref{fig:fit2TeVdown1} does not depend on the mass of the vector-like doublet,
although information on the mass and the couplings resides in the relations 
$U_{RTu}\approx -h_u v_w/M_Q $, $U_{RBd}\approx -h_d v_w/M_Q $. As already mentioned in section \ref{sec:CAA} if we require $h_{u},h_{d}\lesssim 1$ then the vector-like quark needs to be lighter than $\sim 6~\rm TeV$ in order to account for the anomaly.
 An analogous plot can be obtained in the second quadrant. 

We show the $1\sigma$ and $2\sigma$ (green and light-green regions) confidence intervals of the total fit 
($\chi^2_\text{min}+1$, $\chi^2_\text{min}+4$). Moreover, the constraints at $2\sigma$ CL are displayed for the $Z$ boson decay to hadrons (yellow), the weak charge of the proton (red), the atomic parity violation in $^{133}_{78}Cs$ (orange) and the current LHC limits for $t \to u Z$ (solid magenta) and $t \to u H$ (solid purple). We also exhibit the future prospects from FCC-hh on the decay $t \to u Z$ under the two assumption of $0\%$ (dot-dashed magenta) and $5\%$ (dashed magenta) systematic errors~\cite{Liu:2020bem}. The prospects on $t \to u H$ are always subleading and thus are omitted.

In addition, we also perform the fit including only the anomalies, namely eqs. \eqref{eq:det_A}, \eqref{eq:det_B}, \eqref{eq:det_C} and \eqref{eq:mwav} without including other experimental constraints and depict the 1 $\sigma$ region ($\chi^2_\text{min}+1$) in the plots (blue bands). In particular,
the mixing with the first generation
$V_{R ud}=U_{RTu}^*U_{RBd} =  -0.68(27) \times 10^{-3}$ corrects the unitarity relation (see figure \ref{fig:fit2TeVdown1}). As it can be inferred from figure \ref{fig:fit2TeVdown2}, the coupling of the vector-like doublet to the top is crucial in resolving the $m_W$ anomaly.
In figure \ref{fig:fit2TeVdown3} we plot the result of the fit for $h_t$ as a function of $M_Q$, marginalizing over the other parameters and imposing $h_u, h_d \leq 1$. It follows that we can have $h_t\lesssim 1$ for vector-like quark masses below $3~\rm TeV$.

\begin{figure}[t]
\centering
\begin{subfigure}{0.49\textwidth}
\includegraphics[width=\textwidth]{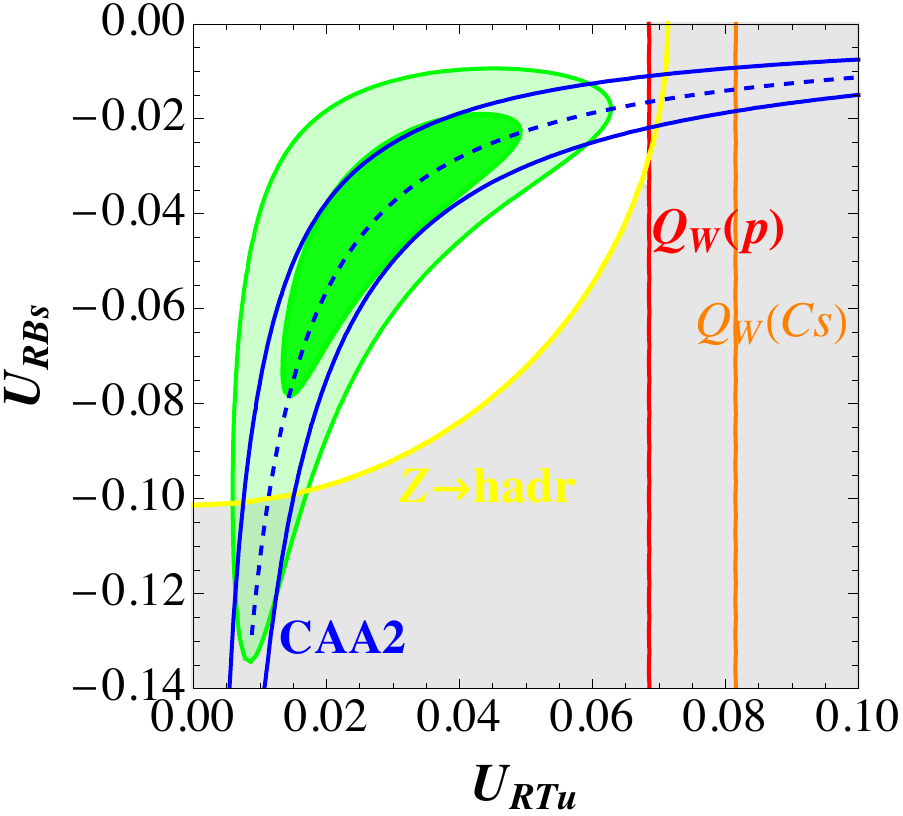}
\caption{\label{fig:fit2TeVstrange1}}
 \end{subfigure}
\hfill
\begin{subfigure}{0.47\textwidth}
\includegraphics[width=\textwidth]{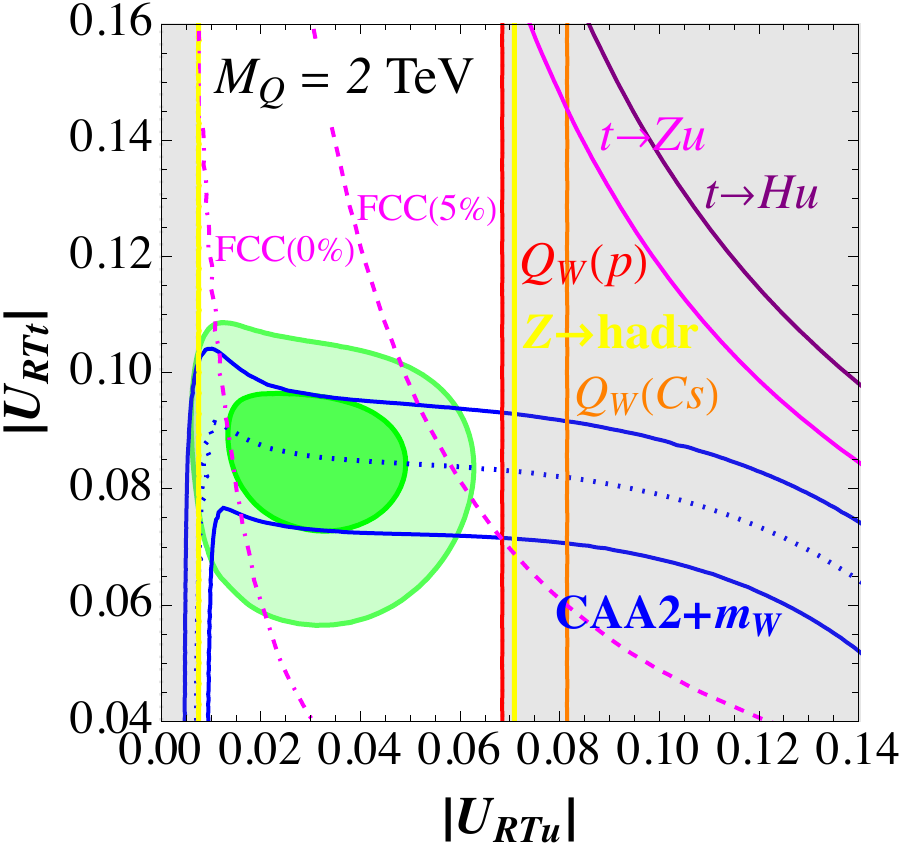}
\caption{\label{fig:fit2TeVstrange2}}
 \end{subfigure}
\caption{\label{fig:fit2TeVstrange} 
Parameter space in the scenario with one vector-like quark doublet coupling with
up, strange and top quarks (see eq. \eqref{eq:vrd}).
$1\sigma$ and $2\sigma$  
preferred regions of 
the mixing parameters are indicated (green and lighter green),
($\chi^2_\text{min}+1$, $\chi^2_\text{min}+4$).
We also show the region excluded at $2\sigma$ by experimental bounds (grey) and
$1\sigma$ interval 
which would explain the CAA2 and the $m_W$-mass shift (blue band), without including other constraints. In figure \ref{fig:fit2TeVstrange2}, $M_Q=2\,\text{TeV}$ is assumed for the mass of the vector-like doublet, and we indicate the experimental bounds on the left side 
using the conservative value $|V_{Rus}|=0.75\times 10^{-3}$ (grey region on the left side).}
\end{figure}

\paragraph{
\textbf{CAA2 + $\bm{m_W}$ ($\bm{h_u,~h_s,~h_t,~M_Q}$):}}
In this scenario the $\chi^2$ fit is performed using the parameters $U_{RTu}$, $U_{RBs}$, $U_{RTt}$, $V_{Lus}$ and including the same observables as before (i.e. determinations of the Cabibbo angle, $m_W$, $\Gamma(Z \to \rm had)$, $Q_W(p)$ and $Q_W(Cs)$). The best-fit point is found to be 
\begin{align}
 U_{RTu}=\pm0.031~, \quad U_{RBs}=\mp0.035~, \quad |U_{RTt}|=0.085~, \quad V_{Lus}=0.22457~,
\end{align}
where again $U_{RTt}$ is determined for $M_Q=2$~TeV. The improvement over the SM is slightly better in this case, $\chi^2_\text{SM}-\chi^2_\text{min}=20.0$, where the remaining discrepancy is due to CAA1. In figures \ref{fig:fit2TeVstrange1} and \ref{fig:fit2TeVstrange2} we illustrate the results of the fit on the $U_{RTu}$-$U_{RBs}$ and $U_{RTu}$-$U_{RTt}$ planes, respectively, marginalizing over the other variables. The $1\sigma$ and $2\sigma$ regions of the fit
($\chi^2_\text{min}+1$, $\chi^2_\text{min}+4$) as well as the current and future constraints are presented with the same colors as in figure \ref{fig:fit2TeVdown}. 
The plot in figure \ref{fig:fit2TeVstrange1} does not change with the mass of the vector-like quark, but from the relations 
$U_{RTu}\approx -h_u v_w/M_Q $, $U_{RBs}\approx -h_d v_w/M_Q $ we arrive as before at the upper bound of
 $M_Q\lesssim 6$~TeV when $h_{u},h_{s}\lesssim 1$.
 
We also perform the stand-alone fit of CAA2 and $m_W$ and exhibit the 1 $\sigma$ region in the plots (blue bands). 
The mixing with up and strange quarks obtained from this fit
$V_{R us}=U_{RTu}^*U_{RBs}\approx h_{u}^\ast {h}_{s}v^2_w/M^2_Q = - 1.13(38) \times 10^{-3}$ (see fig. \ref{fig:fit2TeVstrange1})
induces the difference between the vector coupling from semi-leptonic $K\ell 3$ decays 
and the axial-vector coupling from leptonic $K\mu2/\pi\mu2$ decays.
Furthermore, we infer from \ref{fig:fit2TeVstrange2} that the mixing with the top is also in this case the main source of the modification of the $T$ parameter and $m_W$ prediction. 
 By plotting this coupling against the mass of the vector-like doublet we obtain similar plot as in figure \ref{fig:fit2TeVdown3}, therefore it is omitted. 

Summarizing, one vector-like doublet with mass of few TeV can explain the new measurement of $m_W$ together with 
either the tension between the determinations from kaon decays or the deficit in the CKM unitarity. At high energies a unique prediction of the model is the enhancement of the top decay rate to the up quark and a $Z$ or Higgs boson. Although, the prospects in HL-LHC are only slightly improving over the current bound (and hence are not shown in the plot), it is noteworthy, that most of the parameter space of the model can be explored in the FCC-hh. As a matter of fact, according to the optimistic scenario with respect to the systematics, the whole $1 \sigma$ region will be covered.

\subsection{Vector-like quark singlets}

\label{sec:singlets}

\begin{figure}[t]
\centering
\includegraphics[width=0.46\textwidth]{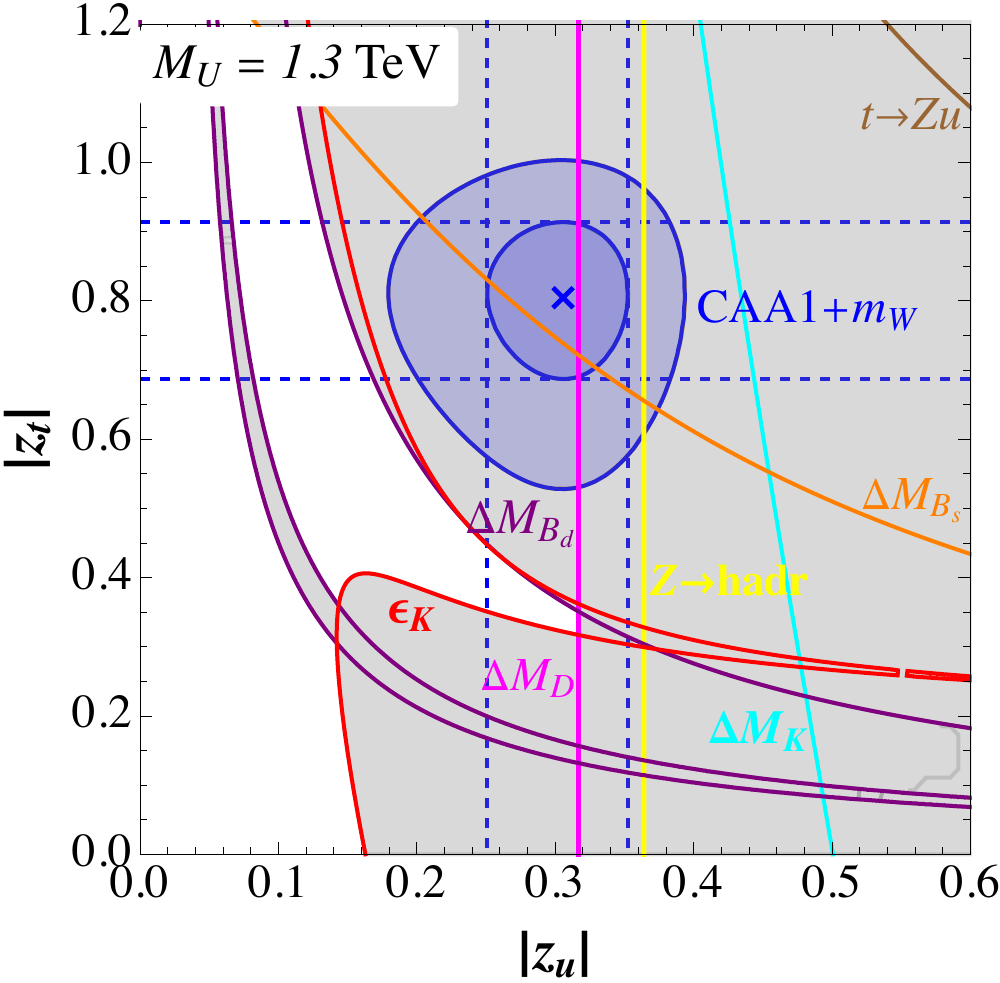} 
\includegraphics[width=0.46\textwidth]{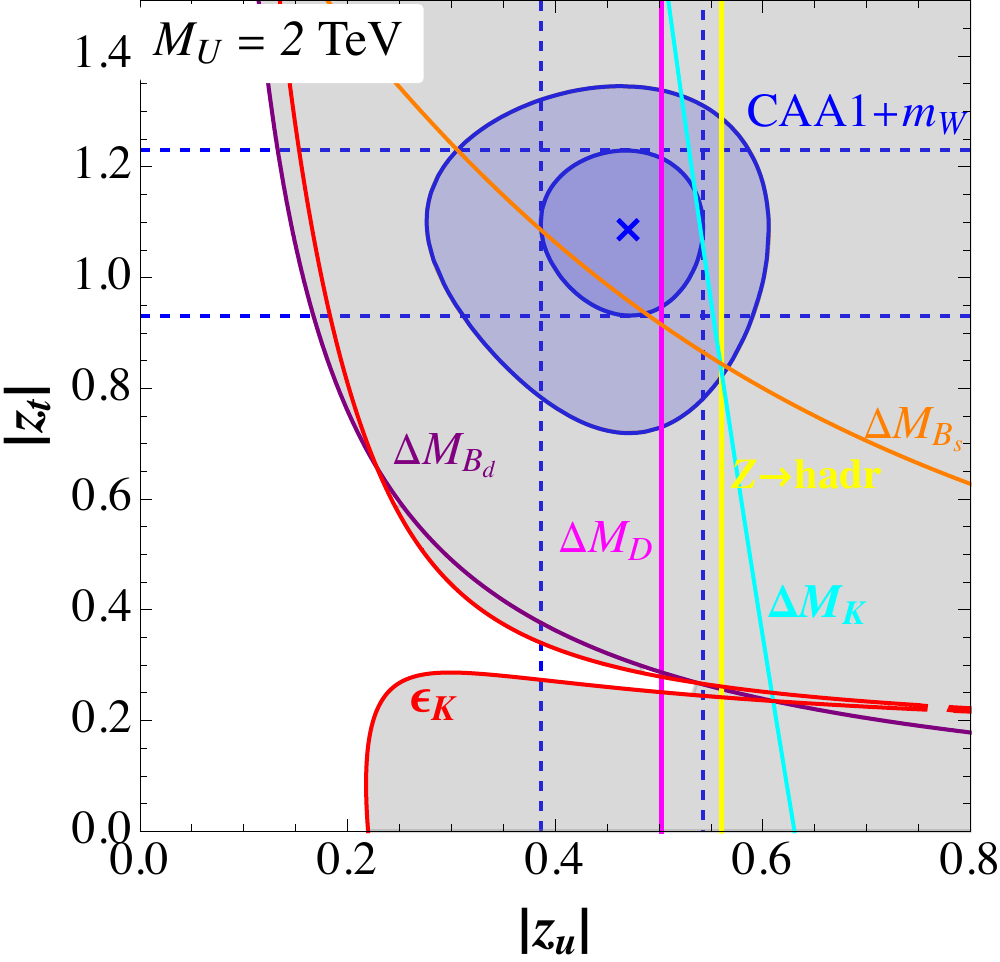} 
\caption{\label{fig:upplot} Parameter space in the scenario with one up-type vector-like singlet for the couplings with up and top quarks, assuming $M=1.3\,\text{TeV}$ (left) and $M=2\,\text{TeV}$ (right)
for the mass of the vector-like singlet.
The $1\sigma$ and $2\sigma$ confidence intervals (blue regions)
of the parameters $z_u$ and $z_t$ obtained from the fit of Cabibbo angle determinations and $m_W$ are shown ($\chi_\text{min}+1$, $\chi_\text{min}+4$).
 We set $U_{LUc}\approx 0.003$,
$\text{Arg}(h_u^*h_c) = -3.0$, and $\text{Arg}(h_u^*h_t) = 2.8$. We see that the region of interest is excluded at $2\sigma$ by experimental bounds (grey).}
\end{figure}

The unitarity anomaly in the first row of CKM is also explained by vector-like quark $SU(2)_L$ singlets, i.e. the down-type $D_{L,R}$ with SM quantum numbers $(\bf{3},\bf{1})_{-1/3}$
or the up-type $U_{L,R}$ with $(\bf{3},\bf{1})_{2/3}$.
The Yukawa couplings and mass terms for the up-type singlet are
\begin{align}
\label{yu}
&\mathcal{L}_{\rm Y} \supset +Y_{uij} \overline{q}_{Li} \tilde{\varphi} u_{Rj}+
z_{i}\tilde{\varphi}\bar{q}_{L i}U_{R}+M_{U}\bar{U}_{L}U_{R}+\text{h.c.}~,
\end{align}
while for the down-type
\begin{align}
\label{yd}
&\mathcal{L}_{\rm Y} \supset +Y_{dij}\overline{q}_{Li} \varphi d_{Rj}+ 
w_{i}\varphi\bar{q}_{Li}D_{R}+M_{D}\bar{D}_{L}D_{R}+\text{h.c.}
\end{align}
where $i,j=1,2,3$ are the family indexes. 
The mass matrices can be diagoanlized via bi-unitary transformations 
$U_{UL}^\dagger {\cal \tilde{M}}_u U_{UR}=\text{diag}(y_uv_w,y_cv_w,y_tv_w,M_{U'})$ and  
analogously for down-type. We can choose the basis in which
the Yukawa submatrix of up-type quarks $\hat{U}_{UL}^\dag Y_u\hat{U}_{UR}$ is diagonal and define
$(z_u,z_c,z_t)= \hat{U}_{UL} z $ 
(and vice versa for down-type).

Using the results in ref.~\cite{Lavoura:1992np} (also reported in the appendix \ref{app:oblique}), the shift of the oblique parameters in the presence of the up-type singlet is given by
\begin{align}
&  T_U \approx \frac{3}{16\pi}\frac{ v_w^2}{s^2c^2\, m_Z^2}\frac{v_w^2}{M_U^2}
\left[ y_t^2|z_t|^2\left(\ln\frac{M_U^2}{m_t^2}-1\right) +
\frac{1}{2}\big( |z_u|^2+|z_c|^2+|z_t|^2  \big)^2
\right]~,  \\
&  S_U \approx \frac{3}{2\pi}
\left[ 
 \sum_{\alpha=u,c,t}\frac{v_w^2}{M_U^2} |z_\alpha|^2 
\left( -\frac{5}{9}+\frac{1}{3}\ln\frac{M^2_U}{m_\alpha^2} \right)
-\frac{1}{9}\sum_{\beta =d,s,b} |V_{U\beta}|^2 \ln\frac{M^2_U}{m_\beta^2} 
\right]~, \\
&  U_U \approx  \frac{3}{2\pi}
\left[ 
\sum_{\alpha=u,c,t}\frac{v_w^2}{M_U^2} |z_\alpha|^2 
\left( \frac{5}{9} -\frac{1}{3} \ln\frac{M^2_U}{m_\alpha^2} \right)
+\frac{1}{3}\sum_{\beta =d,s,b} |V_{U\beta }|^2 \ln\frac{M^2_U}{m_\beta^2} 
\right]~, 
\end{align}
while for the down-type
\begin{align}
&  T_D \approx \frac{3}{16\pi}\frac{ v_w^2}{s^2c^2\, m_Z^2}\frac{v_w^2}{M_D^2}\left[ -y_t^2|w_b|^2\ln\frac{M_D^2}{m_t^2}
+\frac{1}{2}\left( |w_d|^2+|w_s|^2+|w_b|^2  \right)^2 \right]~,
 \\
&  S_D \approx \frac{3}{2\pi}
\left[ 
 \sum_{\beta=d,s,b}\frac{v_w^2}{M_D^2} |w_\beta|^2 
\left( -\frac{5}{9}+\frac{1}{3}\ln\frac{M^2_D}{m_\beta^2} \right)
+\frac{1}{9}\sum_{\alpha=u,c,t} |V_{\alpha D}|^2 \ln\frac{M^2_D}{m_\alpha^2} 
\right]~,  \\
&  U_D \approx  \frac{3}{2\pi}
\left[ 
\sum_{\beta=d,s,b}\frac{v_w^2}{M_D^2} |w_\beta|^2 
\left( \frac{5}{9} -\frac{1}{3} \ln\frac{M^2_D}{m_\beta^2} \right)
+\frac{1}{3}\sum_{\alpha=u,c,t} |V_{\alpha D}|^2 \ln\frac{M^2_D}{m_\alpha^2} 
\right]~.
\end{align}
where $V_{U \beta}$, $V_{ \alpha D}$ are the extra elements of the mixing matrices
in left-handed charged current interactions with $W$ boson in mass basis. 
Then, the $m_W$ prediction is modified according to eq. \eqref{eq:deltamW}.
Also in this case the main contribution comes from the $T$ parameter.
To address the CAA1 with the down-type singlet, a mixing with the first generation $|w_d| v_w/M_D \sim 0.04 $ is needed \cite{Belfatto:2019swo,Belfatto:2021jhf}. However, 
in the allowed range of values for the Yukawa couplings, a positive $m_W$ shift cannot be generated.
In fact, flavor-changing kaon processes impose strict constraints on $w_s$, while the mixing $w_bv_w/M_D$ is bounded by the $Z$ decay rate into hadrons, which gives $|w_b|v_w/M_D<0.03$. In addition, the constraints from flavor-changing $B$-meson decays and neutral $B$-meson systems imply $|w_b^*w_d|v_w^2/M^2_D<2 \times 10^{-4}$. These constraints, combined with limits from the $D^0$ meson system, set an upper limit on the mass of the down-type vector-like quark at around $1.5$~TeV \cite{Belfatto:2021jhf}. Thus, we obtain an approximate upper limit of $|w_b| \lesssim 0.05$, which cannot induce the positive contribution to $m_W$.

In the case of the up-type singlet,
the Lagrangian for the charged-current interaction reads
\begin{align}
\label{ccu}
\mathcal{L}_\text{cc}& = -
\frac{g}{\sqrt{2}} W_\mu^+ \left(\begin{array}{cccc}
\bar{u}_L & \bar{c}_L & \bar{t}_L & \bar{U'}_L
\end{array}\right)\gamma^\mu
V
\left(\begin{array}{c}
d_{L} \\ s_{L} \\ b_{L} 
\end{array}\right)  +\text{h.c.}~,
\end{align}
where $V$ is a $4\times 3$ matrix and the upper left $3 \times 3$ submatrix corresponds to the CKM matrix in the limit of decoupled new physics. The elements of the fourth row are
\begin{align}
V_{U'd}  \approx   \frac{z^{*}_{u}v_w}{M_{U}}~, \quad  V_{U's}\approx \frac{z^{*}_{u}v_w}{M_{U}} V_{us} +\frac{z^{*}_{c}v_w}{M_{U}} V_{cs}+\frac{z^{*}_{t}v_w}{M_{U}}V_{ts}  ~, \quad 
 V_{U'b} \approx \frac{z^{*}_{t}v_w}{M_{U}}~.
\end{align}

For the first row it holds that
\begin{align}
\label{newunup}
& |V_{ud}|^2+|V_{us}|^2+|V_{ub}|^2=
1-|U_{LUu}|^2\approx 1- \frac{|h_{u}|^2v^2_w}{M^2_{U}}~, 
\end{align}
and thus the anomaly can be resolved if the mixing with the first family is $|U_{LUu}| =0.041(7)$.

However, the mixing of SM quarks with the vector-like quarks induces non-standard couplings of the 
$Z$ boson with the left-handed up quarks because of different weak isospin couplings. These couplings are constrained by limits on FCNCs
($\Delta F = 2$ transitions, $B$ and $D$ meson decays)
as well as flavour-conserving processes, e.g. $Z$-boson decay. 
We list the most important constraints in appendix \ref{app:fcnc} 
and a more detailed analysis can be found in refs. \cite{Belfatto:2021jhf,Branco:2021vhs,Botella:2021uxz}.
As a result, although a combination of couplings and relative phases can be found to 
explain the anomalies separately, a simultaneous explanation can be in strong contrast with experimental bounds. 

In fact,
the coupling with the second generation  $z_c$ is strongly constrained by neutral $D$-mesons system. By requiring that new physics contribution on to the mass splitting cannot exceed the experimental value $\Delta M_D=6.56(76) \times 10^{-15}$~GeV \cite{Workman:2022ynf}  at $2\sigma$ CL, one gets the limit
\begin{align}
& |U_{LUu}^*U_{LUc}|\approx |z_u^*z_c|v_w^2/M_U^2 < 1.3 \times 10^{-4} \,\Big[1+(M_U/3.1~\text{TeV})^2\Big]^{-1/2}~.
\label{eq:DDup-text}
\end{align}
However, the coupling with the charm cannot vanish, since it will turn out to be
necessary to interfere with other contributions
in $K^0$-$\bar{K}^0$ system (see eq. \eqref{eq:kkup}) to satisfy constraints on CP violation (see also ref. \cite{Botella:2021uxz}).
Taking into account the bound on $z_c$ from the $D$-mesons mass difference,
$B^0-\bar{B}^0$ mixing gives a bound which is approximately
\begin{align}
&(V_{U'b}^*V_{U'd})^2\frac{1}{4}\frac{M_U^2}{m_w^2} \lesssim  
(V_{tb}^*V_{td}) S_0(m_t^2/m_W^2) \Delta_{B_d}~,
\end{align}
where $S_0(m_t^2/m_W^2)$ is the Inami-Lim function (see eq. \eqref{eq:S0x})  \cite{Inami:1980fz}.
Requiring that the fraction of new physics contribution to the mixing mass is at most $30\%$ of the SM contribution ($\Delta_{B_d}=0.3$) \cite{Bona:2022zhn} one gets
\begin{align}
&|V_{U'b}^*V_{U'd}|\approx |z_u^*z_t| v_w^2/M_U^2 \lesssim 1.1\times 10^{-3} \left[\frac{1\, \text{TeV}}{M_U}\right]~.
\label{eq:BBup-text}
\end{align}
When $|z_u|v_w/M_U =0.04$ is set for CKM unitarity, 
the constraints imposed by $B^0-\bar{B}^0$ system and CP violation in $K^0$-$\bar{K}^0$ system cannot be simultaneously satisfied  by any choice of value and phase of $z_t$, unless $z_c$ is turned on, without violating the constraint from 
$D^0$-$\bar{D}^0$. Assuming the relative phase of $h_u$ and $h_c$ is chosen to 
compensate the other contributions in $\epsilon_K$, 
the limit in eq. \eqref{eq:BBup-text} translates into the bound
$z_t\lesssim 0.15$. 
The necessary $m_W$ enhancement would require instead a coupling as large as $z_t =0.67(10)$ for $M_U=1$ TeV ($z_t =1.08(15)$ for $M_U=2$ TeV).
As a consequence, 
it seems hard to justify both the apparent CKM unitarity deficit and the $m_W$ mass shift. 

In the following, we analyze the $z_u$-$z_t$ parameter space in more detail and illustrate the result in figure \ref{fig:upplot}, 
Limits from LHC exclude up-type vector-like singlets coupling to the top for masses lower than
$~\sim1.3$~TeV \cite{ATLAS:2018ziw}, so we assume
vector-like quark mass of $M_U=1.3\, \text{TeV}$ (left plot) and $M_U=2\, \text{TeV}$ (right plot).
We perform a $\chi^2$ fit of the Cabibbo angle determinations in eqs. \eqref{eq:det_A}, \eqref{eq:det_B}, \eqref{eq:det_C}, and  the $m_W$ mass (\ref{eq:mwav}).
We illustrate 
the $1\sigma$ and $2\sigma$ intervals (blue and lighter blue regions)
of the parameters $z_u$ and $z_t$ obtained from the fit ($\chi_\text{min}+1$, $\chi_\text{min}+4$).
We indicate the constraints from neutral-mesons systems
(see appendix \ref{app:fcnc} for details), i.e. $K^0$-$\bar{K}^0$:
$\epsilon_K$ (red) and $\Delta m_K$ (cyan), 
$B_{d}^0$-$\bar{B}_d^0$ (purple), $B_{s}^0$-$\bar{B}_s^0$ (orange), $D^0$-$\bar{D}^0$ (magenta),
$Z$-boson decay into hadrons (yellow), and $t\rightarrow Zu$ branching ratio (brown).
We fix the other parameters
$|z_c|$, $\text{Arg}(z_u^*z_c)$, $\text{Arg}(z_u^*z_t)$
at convenient values.
In particular, the phase of $z_t$ 
is selected to reduce the contribution in the $B_d^0$-mesons system and the phase and value of the coupling $z_c$
is set in order to compensate the CP-violating effect in $K^0$-$\bar{K}^0$.

As can be seen by the projections, 
allowed regions can be found to solve CAA1 or $m_W$. However,  
the limit set by $B_{d}^0$-$\bar{B}_d^0$ is robust against variations of the other parameters and still excludes the preferred region for the combined explanation.
Moreover, constraints become more stringent with increasing mass and in any case the mass of the up-type vector-like singlet cannot exceed about $\sim 2.5$~TeV \cite{Belfatto:2021jhf} and still accommodate CAA1.

\subsection{Other mediators}

In the following we list the other possible mediators that have been discussed in the context of the Cabibbo angle anomalies and comment on their compatibility with the CDF-II result.

 \begin{enumerate}
     \item[i)] \emph{singly-charged scalar singlet}~\cite{Crivellin:2020oup,Crivellin:2020klg,Marzocca:2021azj}: In this case the solution to CAA1 is based solely on the modification of $G_F$, which is in turn a consequence of the tree-level contribution to the muon decay rate. As already discussed in the introduction, this effect is  are correlated with a negative shift in the $W$-boson mass~\cite{Belfatto:2019swo,Crivellin:2021njn,Cirigliano:2022qdm} and is thus disfavored.
     \item[ii)] \emph{vector-like leptons}~\cite{Crivellin:2020ebi,Endo:2020tkb,Alok:2020jod}: The new physics effects generated by the presence of vector-like leptons modify the $W$ boson couplings and thus the $\beta$ decays directly. However, also in this case, $G_F$ is affected and worsens the tension in $m_W$.
     \item[iii)] \emph{vector boson singlet}~\cite{Buras:2021btx}: This mediator can in principle modify the $m_Z$ via mixing with the $Z$ boson and that could translate into a positive shift in the prediction of $m_W$~\cite{Strumia:2022qkt,Alguero:2022est}, but it is found to be incompatible with CAA1 as a one-particle solution (at least when the flavour-diagonal couplings are non-zero).\footnote{However, in the context of horizontal gauge symmetries, in ref. \cite{Belfatto:2019swo} the shift in the muon decay constant is induced by flavour-changing gauge bosons related to a family symmetry in the left-handed lepton sector.}
     \item[iv)] \emph{vector boson $SU(2)_L$ triplet }~\cite{Capdevila:2020rrl}: The vector triplet can alleviate the tension in CAA1 by modifying the muon decay rate at tree level, but as previously for the scalar singlet, the $W$-boson mass is decreased~\cite{Capdevila:2020rrl}. 
     \item[v)] \emph{vector boson $SU(2)_R$ triplet}~\cite{Grossman:2019bzp,Dekens:2021bro}:   
     The $W_R$ is the only field of this list that can generate right-handed currents necessary to resolve CAA2 (and possibly also the CDF-II anomaly). Nevertheless, the gauge boson needs to be relatively light, a scenario which is excluded in the minimal left-right symmetric model.
    \item[v)] \emph{leptoquakrs}~\cite{Crivellin:2021bkd,Crivellin:2021egp}: These fields can induce tree-level contributions to $\beta$ decays. However, they are excluded not only by flavor-changing low-energy bounds but also by direct searches at colliders.
 \end{enumerate}

\section{Conclusion}
\label{sec:conclusion}

In this paper we have demonstrated that a vector-like quark doublet 
provides a simple extension to the Standard Model to accommodate the Cabibbo angle anomalies and the measurement of the $W$-boson mass by CDF-II. In fact, tree-level mixing with light quarks induces right-handed charged currents, which can be the reason behind the 
former, while the latter is due to the mixing with the top quark, which can produce a sizeable loop-level contribution to the oblique $T$ parameter. The scenario is consistent with the absence of deviations from the Standard Model so far observed in other low- and high-$p_T$ observables. In particular, one generation of vector-like doublet can account for either the violation of CKM unitarity or the reconciliation of the $K\ell3$ and $K\mu2/\pi\mu2$ determinations of $|V_{us}|$, together with the $m_W$ measurement. No fine-tuned cancellations between diagrams are required.

The first scenario is realized when the vector-like doublet predominantly couples to up, down and top quarks.
The unitarity deficit can be induced by a right-handed mixing between up and down quarks
  $ h_{u} {h}_{d} v^2_w/M^2_Q  \sim  - 0.8 \times 10^{-3} $.
 Then, a flavour texture emerges for the Yukawa couplings with the vector-like doublet.
 Couplings with the first generation are of order $h_u \sim h_d \gtrsim 0.3$. 
For the other couplings, it should be $h_s \lesssim \mathcal{O}(10^{-3})$, $h_c\lesssim \mathcal{O}(10^{-2})$, and $ h_b\lesssim \mathcal{O}(10^{-1})$ in order to comply with the stringent constraints from flavour-changing phenomena. Finally, mixing with the top quark is less constrained and $h_t\approx 1$ suffices to generate the $m_W$ anomaly for a mass of $M_Q=2$~TeV. We notice that couplings  of that size between the Higgs and TeV scale fermions can considerably reduce the instability scale~\cite{Gopalakrishna:2018uxn} and potentially provide an argument for the dynamical selection of the electroweak scale~\cite{Khoury:2021zao}.

In the second scenario, a vector-like doublet couples to the up, strange and top quarks. The right-handed mixing
$h_{u} {h}_{s}v^2_w/M^2_Q \sim - 1.3 \times 10^{-3} $ suitably modifies the vector and axial-vector couplings, and thus, the determinations obtained from semi-leptonic and leptonic decays, respectively. 
A similar texture for the Yukawa couplings is required for this scenario, with $h_u \sim h_s \approx 0.3$, $h_t\approx 1$ and similar suppressions for the other couplings as previously.
In both cases, the mass of the vector-like quark should be in the few TeV range, namely, for $|h_t|\lesssim 1$ it must be $M_Q\lesssim 3~\rm TeV$, potentially making it accessible by direct searches in future colliders.

All three discrepancies can be addressed if the fermion sector would be extended by two generations of vector-like doublets that each couples to the up, the top, and either the down or the strange. Alternatively, one can envision a non-minimal scenario in which the first Cabibbo angle anomaly is resolved by some other mechanism and the second by a vector-like doublet coupling predominantly to the second and the third generation. Additionally, the vector-like doublet would not only induce a positive shift in $m_W$, but also compensate for the adverse effect of a modification of the Fermi constant possibly induced by the other mechanism. This is a unique feature of the model featuring the vector-like doublet not shared by other one-particle mediator models.

In order to settle the CKM unitary puzzle, improved experimental inputs for neutron decay time~\cite{Ezhov:2018cta,Callahan:2018iud}, $g_A$ parameter~\cite{Fry:2018kvq,Soldner:2018ycf,Wang:2019pts}, and pion $\beta$ decay~\cite{Czarnecki:2019iwz,PIONEER:2022alm} are expected in the foreseeable future. The leading hadronic uncertainties at both super-allowed $0^+-0^+$ and neutron decays can be reduced by lattice QCD calculations, which improve the estimation of $\gamma W$ box diagrams (and are executable  with the state of the art techniques)~\cite{Seng:2019plg}. Furthermore, $K_{\ell 3}$ decays can be measured at experiments such as LHCb~\cite{AlvesJunior:2018ldo}. For the $W$-boson mass anomaly on the other hand, a confirmation of the CDF-II result from the LHC experiments would be of utmost importance.

Finally, disentangling the new physics contributions specifically due to the vector-like quark can become feasible both at the high-precision as well as the high-intensity frontiers. For example, the measurement of $K_{\mu 3}/K_{\mu 2}$ at NA62 proposed in ref.~\cite{Cirigliano:2022yyo} can distinguish the presence of right-handed charged currents involving strange quarks predicted in this model. Another probe at low-energies can be offered by the P2~\cite{Becker:2018ggl} and MOLLER~\cite{MOLLER:2014iki} experiments, which will perform precise measurements of the Weinberg angle and that would imply improvement of the bounds from atomic parity violation~\cite{Cadeddu:2021dqx}. On the other hand, future colliders can potentially offer the possibility of testing the model at high energies. The smoking-gun signature of the model is the channel $t \to uZ$, where FCC-hh is expected to provide sufficient sensitivity. The LEP bounds on $Z$-boson couplings can also be significantly improved at future $e^+e^-$ colliders such as FCC-ee~\cite{FCC:2018byv}.

\section*{Acknowledgments}

We would like to thank Claudio Andrea Manzari for his valuable help throughout the development of the project
and Antonio Rodríguez Sánchez, Alexander Azatov, Zurab Berezhiani and Vincenzo Cirigliano for useful discussions.
ST is supported by the Swiss National Science Foundation - project n. P500PT\_203156, and by the Center of Theoretical Physics at MIT (MIT-CTP/5538).

\appendix

\section{Bounds from flavour-changing neutral current processes}
\label{app:fcnc}
We summarize here and update
some of the most stringent constraints from flavour-changing phenomena previously
analysed within this framework in ref. \cite{Belfatto:2021jhf}.

\subsection{$K^+\rightarrow \pi^+\nu\bar{\nu}$}
\label{app:Kpimunu}

The decay $K^+\rightarrow \pi^+\nu\bar{\nu}$ is identified as one of the golden modes, since long-distance contributions are negligibly small.
The effective interaction originates from $Z$-penguin and box diagrams 
and it is given by \cite{Buchalla:1995vs}
\begin{align}
\label{kpnnsm}
\mathcal{L}(K\rightarrow \pi\nu\bar{\nu})_\text{SM}& = \-
\frac{4G_F}{\sqrt{2}}\frac{\alpha(M_Z)}{2\pi\sin^2\theta_\text{W}}
\sum_{\ell=e,\mu,\tau}
[V_{cs}^*V_{cd}X^\ell(x_c)+V_{ts}^*V_{td}X^{\ell}(x_t)]
(\overline{s_{L}}\gamma^\mu d_{L})
(\overline{\nu_{\ell L}}\gamma_\mu\nu_{\ell L}) =
\nonumber \\
&=-\frac{4G_F}{\sqrt{2}}\mathcal{F}_{K}(\overline{s_{L}}\gamma^\mu d_{L})\sum_{\ell = e,\mu,\tau}
(\overline{\nu_{\ell L}}\gamma_\mu\nu_{\ell L})~,
\end{align}
$X(x_a)$ are the relevant Inami-Lim function including QCD and electroweak corrections,
 with $x_a=m^2_a/M^2_W$, $a=c,t$.
The experimental measurement for the branching ratio is \cite{Workman:2022ynf}
\begin{align}
\label{kpexp}
\text{Br}(K^+\rightarrow \pi^+\nu\bar{\nu})_\text{exp}=1.14^{+0.40}_{-0.33} \times  10^{-10}~,
\end{align}
which is compatible with the SM prediction $\text{Br}(K^+\rightarrow \pi^+\nu\bar{\nu})_\text{SM}\approx 0.81\cdot 10^{-10}$.
With increasing experimental precision,
any deviation in this channel would point towards new physics.
The mixings with the vector-like doublet induces at tree level the operator
\begin{align}
\label{Lnew}
\mathcal{L}(K\rightarrow \pi\nu\bar{\nu})_{\rm NP}=
\frac{4G_F}{\sqrt{2}}\frac{1}{2}
(U_{RBs}^*U_{RBd})
(\bar{s}_R\gamma^\mu d_R)
\sum_{e,\mu,\tau}(\bar{\nu}_{\ell\text{L}}\gamma_\mu\nu_{\ell\text{L}} )~,
\end{align}
which generates the total branching ratio  
\begin{align}
 &\text{Br}(K^+\rightarrow \pi^+\nu\bar{\nu})  \approx 
 \text{Br}(K^+\rightarrow \pi^+\nu\bar{\nu})_\text{SM} 
  \left|\frac{-\frac{1}{2}U_{RBs}^*U_{RBd}}{\mathcal{F}_{K}}+1\right|^{2}~,
\end{align}
where $\mathcal{F}_{K}\approx (-3.7+i\, 1.1)\times 10^{-6}$ is defined as in ref. \cite{Belfatto:2021jhf}.
We use the experimental limit in eq. \eqref{kpexp} at $2\sigma$ to obtain an upper bound
\begin{align}
\label{kpnncon0}
&\left|-\frac{\frac{1}{2}U_{RBs}^*U_{RBd}}{\mathcal{F}_{K}}+1\right|< 1.5~,
\end{align}
which, depending on the phase $\text{Arg}(U_{RDs}^*U_{RDd})$, 
gives a limit on the couplings with the vector-like doublet:
\begin{align}
& |U_{RBs}^*U_{RBd}|<(0.4 \mbox{---} 2.0 )\times 10^{-5}~.
\end{align}

\subsection{$K_\text{L}\rightarrow \mu^+\mu^-$}
\label{app:Kmumu}

The rare decay $K_\text{L}\rightarrow \mu^+\mu^-$ is a CP-conserving decay and its short-distance contribution is generated by $Z$-mediated penguin and box diagrams.
The effective Lagrangian in the SM
can be written as \cite{Buchalla:1995vs}
\begin{align}
\mathcal{L}(K\rightarrow \mu^+\mu^-)_\text{SM,SD}&=\frac{G_F}{\sqrt{2}}\frac{\alpha(M_Z)}{2\pi\sin^2\theta_\text{W}}
\big(V_{cs}^*V_{cd}Y(x_c)+V_{ts}^*V_{td}Y(x_t)\big)
(\overline{s}\gamma^\mu\gamma_{5} d)(\overline{\mu}\gamma_\mu\gamma_{5}\mu)+\text{h.c.}
= \nonumber \\
&=\frac{G_F}{\sqrt{2}}\mathcal{F}_{L2}(\overline{s}\gamma^\mu\gamma_{5} d)
(\overline{\mu}\gamma_\mu\gamma_{5}\mu) + \text{h.c.}~,
\label{kmumusm}
\end{align}
where $Y(x_a)$, $x_a=m^2_a/M^2_W$,
are the relevant Inami-Lim functions including QCD and electroweak corrections. 
The SM prediction for the short-distance contribution is calculated to be:
$\text{Br}(K_\text{L}\rightarrow \mu^+\mu^-)_\text{SM,SD}\approx 0.9 \times 10^{-9}$  
\cite{Buras:1997fb}. 
However, this decay is dominated by a long-distance contribution from 
a two-photon intermediate state which almost saturates the observed rate 
Br$(K_\text{L}\rightarrow \mu^+\mu^-)_\text{exp}=(6.84\pm 0.11)\times 10^{-9}$ \cite{Workman:2022ynf}. An upper bound on the short distance contribution is estimated in ref. \cite{Isidori:2003ts} as
\begin{align}
\label{eq:SDtot}
&\text{Br}(K_\text{L}\rightarrow \mu^+\mu^-)_\text{SD}<2.5\times 10^{-9}~.
\end{align}
The vector-like doublet induces the decay at tree level
\begin{align}
\label{eq:kmumunew}
\mathcal{L}(K\rightarrow \mu^+\mu^-)_\text{NP}&=
\frac{G_F}{2\sqrt{2}}  U_{RBs}^*U_{RBd}
(\bar{s}\gamma_\mu\gamma_{5} d)
(\bar{\mu}\gamma^\mu\gamma_{5}\mu)+\text{h.c.}
\end{align}
Then we can define the branching ratio given by the amplitude of the short-distance contribution as
\begin{align}
\label{eq:Brkmm}
 &\text{Br}(K_\text{L}\rightarrow \mu^+\mu^-)_\text{SD}\! =
\text{Br}(K_\text{L}\rightarrow \mu^+\mu^-)_\text{SM,SD}\,
  \bigg[1+\frac{\text{Re}(U_{RBs}^*U_{RBd})}{2\text{Re}(\mathcal{F}_{L2})}\bigg]^{2}~,
\end{align}
where $\mathcal{F}_{L2}\approx(-2.1+i\, 0.74)\times 10^{-6}$ is defined as in ref. \cite{Belfatto:2021jhf}.
By using the upper bound in eq. \eqref{eq:SDtot} on the branching ratio
we get
\begin{align}
\label{kmumud}
 &  \left|1+\frac{\text{Re}(U_{RBs}^*U_{RBd})}{2\text{Re}(\mathcal{F}_{L2})}\right| < 1.7~,
 \end{align}
which results in the approximate limit
\begin{align}
-0.3 \times 10^{-5}<\text{Re}(U_{RBs}^*U_{RBd})  <1.1\times 10^{-5}~.
\end{align}

\subsection{$K^0$-$\bar{K}^0$ mixing}
\label{app:K_mixing}

In the SM the short-distance contribution to the transition 
$K^0\leftrightarrow \bar{K}^0$
arises from weak box diagrams.
The two relevant observables describing the mixing are the mass splitting $\Delta M_K=m_{K_L}-m_{K_S}$
and the CP-violating parameter $\epsilon_K$. They are primarily described by
the off-diagonal term $M_{12}$ of the mass matrix of neutral kaons, 
$M_{12}^K=- \langle K^0|\mathcal{L}_{\Delta S=2}|\bar{K}^0\rangle /( 2m_{K^{0}})$, which in the SM is  \cite{Branco:1999fs}
\begin{align}
  M_{12, \text{SM}}^K =
m_{K^{0}} f^2_K \hat{B}_K \frac{G^2_F m^2_W}{12\pi^2} &\big(\eta_1 (V_{cs}V_{cd}^*)^{2}  S_0(x_c)+\eta_2  (V_{ts}V_{td}^*)^{2}  S_0(x_t)+ \notag \\ &\quad +
2\eta_3 (V_{cs}V_{cd}^*)(V_{ts}V_{td}^*) S_0(x_c,x_t)\big)~,
\label{eq:M12K_SM}
\end{align}
where $x_a=m_a^2/m_W^2$,
$f_K$ is the kaon decay constant, which can be estimated in lattice QCD to be
$f_K=155.7(0.7)$ MeV
\cite{FlavourLatticeAveragingGroupFLAG:2021npn},
$m_{K^{0}}=497.611\pm0.013$~MeV is the neutral kaon mass and the factors
$\eta_1 = 1.87\pm 0.76 $ \cite{Brod:2011ty}, $ \eta_2 = 0.5765 \pm 0.0065$ \cite{Buras:1990fn} and 
$\eta_3 = 0.496\pm 0.047 $ \cite{Brod:2010mj} 
describe short-distance QCD effects. 
The factor $\hat{B}_{K}$ is the correction to the vacuum insertion approximation, which is calculated
lattice QCD 
$\hat{B}_K=0.7625(97)$ \cite{FlavourLatticeAveragingGroupFLAG:2021npn}.
The Inami-Lim functions are \cite{Inami:1980fz}
\begin{align}\label{eq:S0x}
 S_0(x)=x & \left( \frac{4-11x+x^2}{4(1-x)^2}-\frac{3x^2\ln x}{2(1-x)^3} \right)
~, \\
 S_0(x_j,x_k)=x_jx_k &\left[\left(\frac{1}{4}-\frac{3}{2(x_j-1)}-\frac{3}{4(x_j-1)^2}\right)
\frac{\log x_j}{x_j-x_k}+ (x_j\leftrightarrow x_k)
-\frac{3}{4(x_j-1)(x_k-1)}\right]~.
\label{eq:S0xy}
\end{align}
The modulus and the imaginary part of the mixing mass $M^K_{12}$
describe short-distance contributions in the mass splitting and CP-violation in $\bar{K}^0\leftrightarrow K^0$ transitions
\cite{Buchalla:1995vs}
\begin{align}
&\Delta M_K \approx 2|M^K_{12}|
+\Delta m_{K,\text{LD}} 
\, , \qquad
 |\epsilon_K| \approx \frac{|\text{Im} M_{12}^K|}{\sqrt{2}\Delta M_K}~,
\end{align}
(using the phase choice $CP|K^{0}\rangle =-|\bar{K}^{0}\rangle$, in the standard parameterization of $V_\text{CKM}$).
$\Delta m_{K,\text{LD}}$ is the long-distance contribution which is difficult to evaluate \cite{Bai:2014cva,Bai:2018mdv}.
However the short distance contribution 
gives the dominant contribution to
the experimental determinations \cite{Workman:2022ynf}
\begin{align}
&\Delta M_{K,\rm exp}=(3.484\pm 0.006) \times 10^{-15} \,\text{GeV}  
\, , \qquad
 |\epsilon_{K}|_{\rm exp}=(2.228 \pm 0.011) \times 10^{-3}~.
\end{align}

The new contribution from the vector-like doublet to the mixing mass term of neutral mesons systems includes relevant right-handed currents at both tree and loop level but also chirality-mixing enhanced contributions at loop level. It is given by
\begin{align}
 M_{12,\rm NP}^K \approx &
\frac{1}{3}m_{K^{0}}f_{K}^{2} \, 0.43\, \bigg\{
\frac{G_F}{\sqrt{2}}(U_{RBd}^*U_{RBs})^2 + \frac{G_F^2}{4\pi^2}   \bigg[  
\,\frac{1}{2} M_Q^2 (U_{RBd}^*U_{RBs})^2 +  \nonumber \\ &
-  3.1 \frac{m_{K^0}^2}{(m_d+m_s)^2}  
 \: 
(U_{RBd}^*U_{RBs}) \,
(V_{Ltd}^{*}V_{Lts}) \, m_W^2 f(M^2_Q/m^2_W,m_{t}^{2}/m_W^2)
\bigg]\bigg\}~,
\end{align}
where  $ f(x_Q,x_t)\approx x_t \ln(x_Q)\,/\,4 $
and we used the numerical coefficients calculated in refs. \cite{Garron:2016mva,Buras:2001ra}. 
Bounds on the new physics contribution can be estimated as $|M_{12,\rm NP}^K|<|M_{12, \text{SM}}^K|\,\Delta_{K} $,
$ |\text{Im}M_{12,NP}^K|<|\text{Im}M_{12, \text{SM}}^K| \, \Delta_{\epsilon_{K}} $.
Setting $\Delta_{K}=1$ and using the results in ref. \cite{Bona:2022zhn} at $95\%$ CL, 
(which approximately corresponds to $\Delta_{\epsilon_{K}}= 0.3$) 
we obtain 
\begin{align}
&  |U_{RBs}^{*}U_{RBd}|
 <  6 \times 10^{-7} \mbox{---} 4\times 10^{-4}~,
 \label{eq:vdshhkk}
\end{align}
depending on the relative phase of the couplings and on the mass of the heavy doublet.
In fact, the limit in \eqref{eq:vdshhkk} is computed for $M_Q\approx 2$~TeV, but the limit on the mixing elements 
in \eqref{eq:vrd} $ |U_{RDs}^{*}U_{RDd}|$ becomes stronger with 
increasing mass $M_Q$ \cite{Belfatto:2021jhf}.

In the scenario with extra up-type quark, box diagrams with 
$U'$ quark running in the loop give the contribution
\begin{align}
 M_{12,\rm NP}^K = \frac{1}{3}m_{K^{0}}f_{K}^{2}  \frac{G^2_F m^2_W}{4\pi^2} \, 0.43\,
& \Big((V_{U's}^*V_{U'd})^2 S_0(x_{U'})+
2(V_{U's}^*V_{U'd})(V_{cs}^*V_{cd})S_0(x_c,x_{U'})+  \nonumber \\
&\quad +2(V_{U's}^*V_{U'd})(V_{ts}^*V_{td})S_0(x_t,x_{U'})\Big)~,
\label{eq:kkup}
\end{align}
with the same definitions as before.

\subsection{Neutral $B$ mesons}
\label{app:B_mixing}

In neutral $B$-mesons system long distance contributions are estimated to be small.  
The dominant short-distance contribution to the $B^0_{d}$-$\bar{B}_{d}^0$ mixing
in the SM is given by
\begin{align}
\label{eq:dmBB}
&\Delta M_{B_{d},\text{SM}}=2|M_{12,\text{SM}}^B|=m_{B_{d}}f^2_{B_{d}}B_{B_{d}}\eta_B\frac{G_F^2m_W^2}{6\pi^2}|(V_{tb}V_{td}^*)^2|S_{0}(x_t)~,
\end{align}
where $M_{12,\text{SM}}^{B_d}=- \langle B_d^0|\mathcal{L}_{\Delta B_d=2}|\bar{B}_d^0\rangle /( 2m_{B_d^{0}})$ 
$\eta_B$ is the QCD factor $\eta_B=0.551$ \cite{Buchalla:1995vs}, and 
$B_{B_{d}}$ is the correction factor to the vacuum-insertion approximation. Analogously, the expression for
$B^0_{s}$-$\bar{B}_{s}^0$ system can be obtained by substituting $d \to s$.
 Lattice QCD calculations yield
$f_{B_d}\sqrt{\hat{B}_{B_d}}=210.6(5.5)$~MeV and $f_{B_s}\sqrt{\hat{B}_{B_s}}=256.1(5.7)$~MeV
\cite{FlavourLatticeAveragingGroupFLAG:2021npn}.
The experimental 
result is \cite{Workman:2022ynf}
\begin{align}
&\Delta M_{B_d,\text{exp}}=(3.334\pm 0.013) \times 10^{-13} \, \text{GeV}~, \qquad
\Delta M_{B_s,\text{exp}}=(1.1693\pm 0.0004) \times 10^{-11} \, \text{GeV} ~. 
\end{align}
The additional contribution due to the presence of the vector-like doublet is
\begin{align}
M_{12,\rm NP}^{B_d}\approx &\frac{1}{3} m_{B_{d}} f_{B_{d}}^2\, 0.80\, \bigg\{
\frac{G_F}{\sqrt{2}}(U_{RBd}^*U_{RBb})^2 + \frac{G_F^2}{4\pi^2}   \bigg[  
\,\frac{1}{2} M_Q^2 (U_{RBd}^*U_{RBb})^2 +  \nonumber \\ &
-  3.37  (U_{RBd}^*U_{RBb})  
(V_{Ltd}^{*}V_{Ltb}) \, m_W^2 f(M^2_Q/m^2_W,m_{t}^{2}/m_W^2)  
\bigg]\bigg\}~
\end{align}
where  $ f(x_Q,x_t)\approx x_t \ln(x_Q)\,/\,4 $, $f_{B_{d}}=190.0(1.3)$~MeV \cite{FlavourLatticeAveragingGroupFLAG:2021npn} and we used the numerical results in refs. \cite{FermilabLattice:2016ipl,Buras:2001ra},
and similarly for $B_s$
\begin{align}
M_{12,\rm NP}^{B_s}\approx &\frac{1}{3} m_{B_{s}} f_{B_{s}}^2\,0.79\, \bigg\{
\frac{G_F}{\sqrt{2}}(U_{RBs}^*U_{RBb})^2 + \frac{G_F^2}{4\pi^2}   \bigg[  
\,\frac{1}{2} M_Q^2 (U_{RBs}^*U_{RBb})^2 +  \nonumber \\ &
-  3.14  (U_{RBs}^*U_{RBb})  
(V_{Lts}^{*}V_{Ltb}) \, m_W^2 f(M^2_Q/m^2_W,m_{t}^{2}/m_W^2)  
\bigg]\bigg\}~,
\end{align}
with $f_{B_{s}}=230.3(1.3)$~MeV \cite{FlavourLatticeAveragingGroupFLAG:2021npn} and using the results in refs. \cite{FermilabLattice:2016ipl,Buras:2001ra}.
We can use the constraints obtained in ref. \cite{Bona:2022zhn} at $95\%$ CL
in order to limit the contribution of new physics. 
These bounds approximately yield a constraint
$\Delta M_{B_{d(s)}}^\text{new}< \Delta M_{B_{d(s)}}^\text{SM} \Delta_{B_{d(s)}}$ 
with $\Delta_{B_d}=0.3$, $\Delta_{B_s}= 0.2$.
For $M_q\approx 2$~TeV we obtain
$|U_{RBd}^*U_{RBb}|\lesssim (1.6 \mbox{---} 3.9) \times 10^{-4}$ and $|U_{RBs}^*U_{RBb}|\lesssim (0.6 \mbox{---} 1.6) \times 10^{-3} $,
which however is stronger for $M>2$~TeV. 

Also flavour-changing $B$ decays give a limit on the product $|U_{RBd(s)}^*U_{RBb}|$. 
In particular,
the most constraining processes are the decays $B^0_{d(s)}\rightarrow\mu^+\mu^-$, for which the experimental upper limits are
$\text{Br}(B^0_d\rightarrow\mu^+\mu^-)_\text{exp}<2.1\times 10^{-10}$ at $ 95\%\, \text{CL} $ \cite{Aaboud:2018mst} and
$\text{Br}(B^0_s\rightarrow\mu^+\mu^-)_\text{exp}=(3.01\pm 0.35)\times 10^{-9} $ \cite{Workman:2022ynf}.
We obtain the following bounds $|U_{RBd}^*U_{RBb}|< (0.4 \mbox{---} 2.0) \times 10^{-4} $, 
$|U_{RBs}^*U_{RBb}|< 8 \times 10^{-7} \mbox{---} 8.3 \times 10^{-4} $, with the range determined by the phase.

By combining the results above, we can get the approximate bounds
\begin{align}
&|U_{RBd}^*U_{RBb}|< (0.4 \mbox{---} 1.7) \times 10^{-4}~, \notag \\
&|U_{RBs}^*U_{RBb}|< 8 \times 10^{-7}  \mbox{---} 6.4 \times 10^{-4}~.
\end{align}

In the scenario with vector-like up-type quark, the extra contribution is
\begin{align}
M_{12,\rm NP}^{B_s}\approx\frac{1}{3} m_{B_{d}} f_{B_{d}}^2\frac{G^2_F m^2_W}{4\pi^2}\, 0.80\,&
\left((V_{U'b}^*V_{U'd})^2 S_{0}(x_{U'})+
2(V_{U'b}^*V_{U'd})(V_{tb}^*V_{td})S_{0}(x_t,x_{U'})\right)~,
\label{bbup}
\end{align}
and analogously for $d\rightarrow s$.

\subsection{$D^0$-$\bar{D}^0$ mixing}
\label{app:D_mixing}

The measured value of the mass difference in $D^0$-$\bar{D}^0$ system
is \cite{Workman:2022ynf}
\begin{align}
\label{eq:dmdexp}
&\Delta M_{D,\text{exp}}=(6.6\pm 0.8)\times 10^{-15}~\text{GeV}~.
\end{align}
The short distance contribution in the SM from box and penguin diagrams is estimated to contribute in very small amount
$|M_{12}^D| \sim 10^{-17}\mbox{---} 10^{-16}$ GeV \cite{Branco:1999fs,Petrov:1997fw}.
Long-distance effects
are expected to be large but they are difficult to compute \cite{Branco:1999fs,Cheng:2010rv}. 
Then, new physics can be the main source of the mass difference $\Delta M_{D}$ in the $D^{0}-\bar{D}^0$ system.

The mixing mass induced by the vector-like doublet reads 
\begin{align}
 M_{12,\rm NP}^{D} \approx 
  &\frac{1}{3}f_D^2M_{D^0}\frac{G_{F}}{\sqrt{2}}\big(U^*_{RTu}U_{RTc}\big)^{2}
 \left(1+\frac{G_{F}M_Q^{2}}{4\sqrt{2}\pi^{2}}\right)~,
\end{align}
where $f_D=212.0(0.7)$~MeV \cite{FlavourLatticeAveragingGroupFLAG:2021npn} and left-right contributions are subdominant.
New physics can be the dominant contribution to the mass difference $\Delta M_{D}$ in the $D^{0}-\bar{D}^0$ system.
Then, we can obtain an estimate by requiring the contribution 
$\Delta M_{D\, \text{new}}=2|M_{12,\rm NP}^{D}|$ does not exceed the experimental value in eq. \eqref{eq:dmdexp} at $2\sigma$
\begin{align}
&|U^*_{RTu}U_{RTc}|< 1.0 \times 10^{-4} \left[\frac{f_{Q}(2\,\text{TeV})}{f_{Q}(M_Q)} \right]^{1/2}~,
\end{align}
where, as in ref. \cite{Belfatto:2021jhf}
\begin{align}
& f_{Q}(M)\approx 1+\left(\frac{M}{2.2\,\text{TeV}}\right)^{2}~.
\end{align}

As regards the contribution of an up-type vector-like singlet, we have 
\begin{align}
 M_{12,\rm NP}^{D} & \approx 
\frac{1}{3}f_D^2M_{D^0}
\frac{G_F}{\sqrt{2}}(U_{LUu}^*U_{LUc})^2\left(1+\frac{G_{F}M^{2}_{t'}}{8\sqrt{2}\pi^{2}}  \right)
\approx \nonumber \\ &\approx
\frac{1}{3}f_D^2M_{D^0}\frac{G_F}{\sqrt{2}}(U_{LUu}^*U_{LUc})^2 \Big[ 1+\Big(\frac{M_{t'} }{3.1 \, \text{TeV}}\Big)^{2}\Big]~.
 \label{fmtprimo}
\end{align}

\section{Oblique parameters with vector-like quarks}
\label{app:oblique}

The expressions for the oblique-correction parameters in presence of vector-like quarks were 
computed in ref. \cite{Lavoura:1992np}. We report the general expressions of $T$, $S$ and $U$
for the reader's convenience.
The contribution to the parameter $T$ reads
\begin{align}
T&= \frac{3}{16\pi\sin^2\theta_w\cos^2\theta_w}
\Big\{\sum_{\alpha,i}\big[(|V_{L\alpha i}|^2+|V_{R\alpha i}|^2)\theta_+(x_\alpha,x_i)
+2\text{Re}(V_{L\alpha i}V_{R\alpha i}^*)\theta_-(x_\alpha,x_i)\big] \Big. - \nonumber \\ \Big.
&-\sum_{\beta<\alpha}\big[(|K_{uL\alpha \beta}|^2+|K_{uR\alpha \beta}|^2)\theta_+(x_\alpha,x_\beta)
+2\text{Re}(K_{uL\alpha \beta}K_{uR\alpha \beta}^*)\theta_-(x_\alpha,x_\beta)\big] \Big.- \nonumber \\ \Big.
&-\sum_{j<i}\big[(|K_{dL ij}|^2+|K_{dRij}|^2)\theta_+(x_i,x_j)
+2\text{Re}(K_{dL ij}K_{dRij}^*)\theta_-(x_i,x_j)\big]
\Big\}~,
\end{align}
where
$x_i=m_i^2/m_Z^2$, $K_{d(u)L}$ are defined analogously to $K_{d(u)R}$ in eq. (\ref{eq:Lagr_Z}).
In this appendix we adopt the same convention of ref. \cite{Lavoura:1992np} of using greek letters to denote up-type quarks and latin ones to denote down-type.
The functions $\theta_{+/-}$ are
\begin{align}
&\theta_+(x_i,x_j)=x_i+x_j-\frac{2x_ix_j}{x_i-x_j}\ln\frac{x_i}{x_j}~, \nonumber \\
&\theta_-(x_i,x_j)=2\sqrt{x_ix_j}\Big(\frac{x_i+x_j}{x_i-x_j}\ln\frac{x_i}{x_j}-2\Big)~.
\end{align}
These functions are symmetric under interchange of the arguments and they are zero for $x_i=x_j$.
In the limit $x_i\gg x_j$, it holds that $\theta_+(x_i,x_j)\rightarrow x_i$, $\theta_-(x_i,x_j)\rightarrow 0$.
In our scenario with the vector-like doublet, the matrices $K_{u(d)L}$ correspond to the $4\times 4$ identity matrix,
while $V_L$, $V_R$ and $K_{d(u)R}$ are described in eqs. \eqref{eq:vckmL}, \eqref{eq:vckmR} and \eqref{eq:vnc}, respectively.
The mass splitting is given in eq. \eqref{eq:Msplit}.
The matrices are defined in analogous way for vector-like singlets, in which case
$V_R=0$, $K_{u(d)R}=\mathbf{1}$.

The general result for the contribution to the parameter $U$ is \cite{Lavoura:1992np}
\begin{align}
U= -\frac{N_c}{2\pi}&
\Big\{\sum_{\alpha,i}\big[(|V_{L\alpha i}|^2+|V_{R\alpha i}|^2)\chi_+(x_\alpha,x_i)
+2\text{Re}(V_{L\alpha i}V_{R\alpha i}^*)\chi_-(x_\alpha,x_i)\big] \Big. \nonumber \\ \Big.
&-\sum_{\beta<\alpha}\big[(|K_{uL\alpha \beta}|^2+|K_{uR\alpha \beta}|^2)\chi_+(x_\alpha,x_\beta)
+2\text{Re}(K_{uL\alpha \beta}K_{uR\alpha \beta}^*)\chi_-(x_\alpha,x_\beta)\big] \Big. \nonumber \\ \Big.
&-\sum_{j<i}\big[(|K_{dL ij}|^2+|K_{dRij}|^2)\chi_+(x_i,x_j)
+2\text{Re}(K_{dL ij}K_{dRij}^*)\chi_-(x_i,x_j)\big]
\Big\}~,
\end{align}
where
\begin{align}
&\chi_+(x_i,x_j)=\frac{5(x_i^2+x_j^2)-22x_ix_j}{9(x_i-x_j)^2}+\frac{3x_ix_j(x_i+x_j)-x_i^3-x_j^3}{3(x_i-x_j)^3}\ln\frac{x_i}{x_j}~,
 \nonumber \\
&\chi_-(x_i,x_j)=-\sqrt{x_ix_j}\Big(\frac{x_i+x_j}{6x_ix_j}-\frac{x_i+x_j}{(x_i-x_j)^2}+\frac{2x_ix_j}{(x_i-x_j)^3}\ln\frac{x_i}{x_j}\Big)~.
\end{align}
Also, in this case the functions are symmetric under interchange of the variables and $\chi_\pm(x,x)=0$.
As regard the parameter $S$, the result is \cite{Lavoura:1992np}
\begin{align}
S= -\frac{N_c}{2\pi}&
\Big\{\sum_{\alpha,i}\big[(|V_{L\alpha i}|^2+|V_{R\alpha i}|^2)\psi_+(x_\alpha,x_i)
+2\text{Re}(V_{L\alpha i}V_{R\alpha i}^*)\psi_-(x_\alpha,x_i)\big] \Big. \nonumber \\ \Big.
&-\sum_{\beta<\alpha}\big[(|K_{uL\alpha \beta}|^2+|K_{uR\alpha \beta}|^2)\chi_+(x_\alpha,x_\beta)
+2\text{Re}(K_{uL\alpha \beta}K_{uR\alpha \beta}^*)\chi_-(x_\alpha,x_\beta)\big] \Big. \nonumber \\ \Big.
&-\sum_{j<i}\big[(|K_{dL ij}|^2+|K_{dRij}|^2)\chi_+(x_i,x_j)
+2\text{Re}(K_{dL ij}K_{dRij}^*)\chi_-(x_i,x_j)\big]
\Big\}~,
\end{align}
where
\begin{align}
&\psi_+(x_\alpha,x_i)=\frac{1}{3}-\frac{1}{9}\ln\frac{x_\alpha}{x_i}~, \qquad
&\psi_-(x_\alpha,x_i)=-\frac{x_\alpha+x_i}{6\sqrt{x_\alpha x_i}}~.
\end{align}
These functions do not vanish for $x_\alpha=x_i$, but $\psi_+(x,x)=-\psi_-(x,x)$ and
$\psi_+(x_\alpha,x_i)$ is not symmetric under interchange of the arguments.

\bibliographystyle{JHEP}

\end{document}